\documentclass[11pt]{article} 

\makeatletter\renewcommand{\section}{\@startsection
	{section}{1}{\z@}{-3.5ex plus -1ex minus
		-.2ex}{2.3ex plus .2ex}{\bf }}

\makeatletter\renewcommand{\subsection}{\@startsection{subsection}{2}{\z@}{-3.25ex
		plus -1ex minus
		-.2ex}{1.5ex plus .2ex}{\it }}
\makeatletter\renewcommand{\subsubsection}{\@startsection{subsubsection}{3}{-2.45ex}{-3.25ex
		plus -1ex minus -.2ex}{1.5ex plus .2ex}{\it }}

\makeatletter \@addtoreset{equation}{section}

\usepackage{graphicx}
\graphicspath{{figures/}}
\usepackage{float} 
\usepackage{epsfig} 
\usepackage{verbatim}
\usepackage{hyperref}
\usepackage{easyfig}
\usepackage{bbm}
\usepackage{amsmath}
\usepackage{amsfonts}
\usepackage[compact]{titlesec}
\usepackage{subdepth}
\usepackage{dsfont}
\usepackage{braket}
\usepackage{tikz}
\usepackage[utf8]{inputenc}
\usepackage[T1]{fontenc}
\usepackage{amsmath}
\usepackage[toc]{glossaries}

\usepackage{caption}
\usepackage{subcaption}

\renewenvironment{thebibliography}[1]
{\baselineskip=16pt plus 2pt minus 1pt
	\section*{\large\refname
		\@mkboth{\MakeUppercase\refname}{\MakeUppercase\refname}}%
	\list{\@biblabel{\@arabic\c@enumiv}}%
	{\settowidth\labelwidth{\@biblabel{#1}}%
		\leftmargin\labelwidth
		\advance\leftmargin\labelsep
		\@openbib@code
		\usecounter{enumiv}%
		\let\p@enumiv\@empty
		\renewcommand\theenumiv{\@arabic\c@enumiv}}%
	\sloppy
	\clubpenalty4000
	\@clubpenalty \clubpenalty
	\widowpenalty4000%
	\sfcode`\.\@m}

\let\fn\footnote
\renewcommand{\footnote}[1]{\linespread{1.1}\fn{#1}\linespread{1.29}}

\usepackage{physics}
\usepackage{amsfonts}
\usepackage{amssymb}
\usepackage{mathrsfs}
\usepackage{amsmath,amssymb}
\usepackage{bm}
\usepackage{caption}
\usepackage{amsmath}
\usepackage{graphicx}
\usepackage{cite}

\usepackage[T1]{fontenc}
\usepackage[utf8]{inputenc}

\def\tyng(#1){\hbox{\tiny$\yng(#1)$}}

\newcommand{\be}{\begin{equation}}
	\newcommand{\ee}{\end{equation}}
\newcommand{\bea}{\begin{array}}
	\newcommand{\ea}{\end{array}}
\newcommand{\beqa}{\begin{eqnarray}}
	\newcommand{\eeqa}{\end{eqnarray}}
\newcommand{\nn}{\nonumber}

\usepackage{pdfpages}

\begin{document}
	
	\fontfamily{bch}\fontsize{11pt}{15pt}\selectfont
	\begin{titlepage}
		\begin{flushright}
			
		\end{flushright}
		
		
		\begin{center}
			{\Large \bf Chaotic Dynamics of the Mass Deformed ABJM Model}\\
			~\\
			
			
		\vskip 3em

\centerline{$ \text{\large{\bf{K. Ba\c{s}kan}}}^{a} \,, \, $ $ \text{\large{\bf{S. K\"{u}rk\c{c}\"{u}o\v{g}lu}}}^{a} \,,\,$ $\text{\large{\bf{C. Ta\c{s}cı}}}^{a,b}$}
\vskip 0.5cm
\centerline{\sl $^a$ Middle East Technical University, Department of Physics,}
\centerline{\sl Dumlupınar Boulevard, 06800, Ankara, Turkey}
\vskip 1em
\centerline{\sl $^b$ The Graduate Center, City University of New York}
\centerline{\sl 365 Fifth Ave, New York, NY 10016, U.S.A.}
\vskip 1em
\begin{tabular}{r l}
	E-mails: 
	&\!\!\!{\fontfamily{cmtt}\fontsize{11pt}{15pt}\selectfont kagan.baskan@metu.edu.tr  }\\
	&\!\!\!{\fontfamily{cmtt}\fontsize{11pt}{15pt}\selectfont kseckin@metu.edu.tr} \\
	&\!\!\!{\fontfamily{cmtt}\fontsize{11pt}{15pt}\selectfont ctasci@gradcenter.cuny.edu}
\end{tabular}
			
		\end{center}
		
		\vskip 5 em
		
		\begin{quote}
			\begin{center}
				{\bf Abstract}
			\end{center}
			
			We explore the chaotic dynamics of the mass-deformed Aharony-Bergman-Jafferis-Maldacena model. To do so, we first perform a dimensional reduction of this model from $2+1$ to $0+1$ dimensions, considering that the fields are spatially uniform. Working in the 't Hooft limit and tracing over ansatz configurations involving fuzzy 2-spheres, which are described in terms of the Gomis–Rodriguez-Gomez–Van  Raamsdonk–Verlinde matrices with collective time dependence, we obtain a family of reduced effective Lagrangians and demonstrate that they have chaotic dynamics by computing the associated Lyapunov exponents. In particular, we focus on how the largest Lyapunov exponent, $\lambda_L$, changes as a function of $E/N^2$. Depending on the structure of the effective potentials, we find either  $\lambda_L \propto (E/N^2)^{1/3}$ or $\lambda_L \propto (E/N^2 - \gamma_N)^{1/3}$, where $\gamma_N(k, \mu)$ are constants determined in terms of the Chern-Simons coupling $k$, the mass $\mu$, and the matrix level $N$. Noting that the classical dynamics  approximates the quantum theory only in the high-temperature regime, we investigate the temperature dependence of the largest Lyapunov exponents and give upper bounds on the temperature above which $\lambda_L$ values comply with the Maldacena-Shenker-Stanford bound, $ \lambda_L \leq 2 \pi T $, and below which it will eventually be not obeyed.
			
			\vskip 1em

		\end{quote}
		
	\end{titlepage}
	
	\setcounter{footnote}{0}
	\pagestyle{plain} \setcounter{page}{2}
	
	\newpage

\section{Introduction}
\label{intro}

Studies on exploring the structure of chaotic dynamics emerging from the matrix quantum mechanics have been continuing with growing interest for quite sometime \cite{Sekino:2008he,  Asplund:2011qj,  Shenker:2013pqa, Gur-Ari:2015rcq, Berenstein:2016zgj, Maldacena:2015waa, Maldacena:2016hyu, Aoki:2015uha, Asano:2015eha, Berkowitz:2016jlq, Buividovich:2017kfk, Buividovich:2018scl, Coskun:2018wmz, Baskan:2019qsb}. Early investigations on the chaotic dynamics of Yang-Mills (YM) gauge theories dates back to the 1980's \cite{Matinyan:1981dj, Savvidy:1982wx, Savvidy:1982jk} and in the context of the Banks-Fischler-Shenker-Susskind (BFSS) model \cite{Banks:1996vh} to the work Arefeva \emph{et al}. \cite{Arefeva:1997oyf}. Recent studies are especially motivated by a result due to Maldacena-Shenker-Stanford (MSS) \cite{Maldacena:2015waa}, which briefly states that, under rather general conditions met by a physical system, the largest Lyapunov exponent (which is a measure of chaos in both classical and quantum mechanical systems) for quantum chaotic dynamics is controlled by a temperature-dependent bound and given by $\lambda_L \leq 2\pi T$. It is conjectured that systems which are holographically dual to the black holes are maximally chaotic, meaning that they saturate this bound. This is already demonstrated for a particular fermionic matrix model, namely, the Sachdev-Ye-Kitaev \cite{Maldacena:2016hyu}  model and expected to be so for other matrix models which have a holographic dual such as the BFSS \cite{Banks:1996vh} model. The latter and the Berenstein-Maldacena-Nastase (BMN) model \cite{Berenstein:2002jq} are supersymmetric $SU(N)$ gauge theories, describing the dynamics of the $N$-coincident $D0$-branes in the flat and spherical backgrounds, respectively, and also appear in the Discrete Light Cone Quantization (DLCQ) of M theory in the flat and the \emph{pp}-wave backgrounds \cite{Banks:1996vh, deWit:1988wri, Itzhaki:1998dd,  Berenstein:2002jq, Dasgupta:2002hx, Ydri:2017ncg,Ydri:2016dmy}. It is well known that the gravity dual of the BFSS model is obtained in the 't Hooft limit and describes a phase in which $D0$-branes form a so-called black brane, i.e., a string theoretical black hole \cite{Ydri:2017ncg,Ydri:2016dmy,Kiritsis:2007zza}. 

Classical dynamics of YM matrix models provide a good approximation of the high-temperature limit of the quantum theory. Although this regime is distinguished from that in which the gravity dual is obtained (i.e., the low-temperature limit), early numerical studies conducted in Refs. \cite{Anagnostopoulos:2007fw, Catterall:2008yz} gave no indication of an occurrence of a phase transition between the low- and high-temperature limits, which makes it quite plausible that some features like fast scrambling \cite{Sekino:2008he} of black holes in the gravity dual and temperature dependence of chaotic dynamics may be retained to a certain extent at the high-temperature limit too. For instance, the numerical results obtained in Ref. \cite{Asplund:2011qj} by exploiting the classical dynamics of the BMN model results in a fast thermalization. In Ref. \cite{Gur-Ari:2015rcq}, classical chaotic dynamics of the BFSS models is studied, and there it is found that the largest Lyapunov exponent is given as $\lambda_L = 0.2924(3) (\lambda_{'t \, Hooft} T)^{1/4}$. Therefore, the MSS bound is not obeyed only at temperatures below the critical temperature $T_c \approx 0.015\,,$ while it remains  parametrically smaller than $2 \pi T$ for $T > T_c$. In Ref. \cite{Baskan:2019qsb}, we have studied chaos in massive deformations of the $SU(N)$ Yang-Mills gauge theories in $0+1$-dimensions, with the same matrix content as that of the bosonic part of the BFSS model, by making use of ansatz configurations involving both fuzzy 2- and 4-spheres. Our numerical results have shown very good agreement with the $\lambda_L \propto (E/N^2)^{1/4}$ -type functional behavior of the largest Lyapunov exponent with energy, which, together with the application of the virial and the equipartition theorems, allowed us to put upper bounds on the critical temperature, $T_c$. Depending on the values of the mass parameters, our estimates for $T_c$ were around twice or about an order of magnitude larger than that obtained for the BFSS model in Ref. \cite{Gur-Ari:2015rcq}. In the present paper, we extend and apply the methods we have developed in Ref. \cite{Baskan:2019qsb} to another interesting gauge theory, namely, the massive deformation of the Aharony-Bergman-Jafferis-Maldacena (ABJM) model \cite{Aharony_2008} \cite{Hosomichi:2008jb,Gomis_2008}. Before focusing our attention in this direction, let us also note that not only the BFSS and the BMN matrix models, but even their subsectors at small values of $N$ are quite nontrivial many-body systems, which escape a complete solution to this day. Nevertheless, the chaotic dynamics of the smallest YM matrix model composed of two $2\times 2$ Hermitian matrices with $SU(2)$ gauge and $SO(2)$ global symmetries has recently explored in Ref. \cite{Berenstein:2016zgj} (see also Refs. \cite{Kabat:1996cu, Kares:2004uk} in this context) with the chaotic phase, corresponding to a toy model for a black hole, being controlled by the angular momentum associated to the rigid $SO(2)$ symmetry. In Ref. \cite{Baskan:2021aap}, two of us explored the minimal Yang-Mills-Chern-Simons matrix model and analyzed the effect of the Chern-Simons (CS) coupling on the chaotic dynamics.

 As is well known, the ABJM model is a $2+1$-dimensional $\mathcal{N}=6$ supersymmetric $SU(N)\cross SU(N)$ CS gauge theory at the CS level $(-k,k)$ \cite{Aharony_2008} and describes the dynamics of $N$ coincident $M2$-branes \cite{Aharony_2008, Nastase:2015wjb}. This model consists of four complex scalar fields $C^I$ ($I:1,2,3,4$), as would be expected due to the eight transverse directions to the $M2$-branes, and four Majorana fermions $\psi^I$ to match the bosonic and fermionic degrees of freedom as a minimal requirement for the presence of supersymmetry.  These fields are coupled bifundamentally to the $SU(N)$ CS gauge fields $A_{\mu}$ and $\hat{A}_\mu$, i.e. they carry the $(N, \bar{N})$ representation of the $SU(N)\cross SU(N)$ group. The model has the $R$-symmetry group $U(1) \times SU(4)$ under which both the complex scalars and the fermions transform in the four-dimensional fundamental representation of $SU(4)$ and carry $+1$ charge under the $U(1)$ factor. The ABJM model is dual to type-IIA string theory on $AdS_4 \times S^7/ \mathbb{Z}^k$ (this becomes $AdS_4 \times  \mathbb{C} P^3$ in the $k \rightarrow \infty$ limit) via the AdS/CFT correspondence \cite{Aharony_2008, Nastase:2015wjb}, and it possesses a massive deformation due to Hosomichi \emph{et al.} \cite{Hosomichi:2008jb} and Gomis \emph{et al.} (GRVV) \cite{Gomis_2008}, preserving all the supersymmetry, but breaking the $R$ symmetry. It is this model on which we focus our attention in the present paper. The vacuum configurations in this model are given by the GRVV matrices, which describe fuzzy 2-spheres as a somewhat intricate analysis demonstrates \cite{Nastase:2015wjb, Gomis_2008}. This feature is similar and comparable to the BMN model, which also has fuzzy spheres as the vacuum solutions. Our aim here is to explore the chaotic dynamics emerging from this model at the classical level as an approximation to the quantum theory in the high-temperature regime using both analytic and numeric techniques and determine upper bounds on the temperature of the system at consecutively higher matrix levels, above which the MSS bound is satisfied and below which it will eventually not be obeyed. The latter is naturally expected to occur, since, as we already noted, the classical treatment of the model could approximate the dynamics of the full quantum theory only at sufficiently high temperatures. Toward this aim, we first perform a dimensional reduction of this model from $2+1$- to $0+1$ dimensions by considering that the fields are spatially uniform. We work in the 't Hooft limit and focus on two distinct ansatz configurations both involving fuzzy 2-spheres, which are described in terms of the GRVV matrices. These configurations have collective time dependence, which are introduced by real functions of time multiplying the latter. Tracing over these configurations yields a family of reduced effective Lagrangians and we demonstrate that they have chaotic dynamics by computing their Lyapunov exponents. In particular, we direct our attention to examine how the largest Lyapunov exponent, $\lambda_L$, changes as a function of $E/N^2$. It turns out that, depending on the structure of the effective potentials, we find that either  $\lambda_L \propto (E/N^2)^{1/3}$ or $\lambda_L \propto (E/N^2 - \gamma_N)^{1/3}$, where $\gamma_N(k, \mu)$ is a constant determined in terms of the CS coupling $k$, the mass $\mu$, and the matrix level $N$. This power-law response of $\lambda_L$ to energy is also anticipated and supported by the exact scaling symmetry possessed by the model in the massless limit as will be discussed in the next section. Making use of our numerical results, and evoking the virial and equipartition theorems, we explore the implications for the aforementioned MSS conjecture in the context of this model. The main outcomes are the upper bounds we obtain on the temperatures above which largest Lyapunov exponents comply with the MSS bound and below which it will eventually be not obeyed. At the same time, we demonstrate that with increasing matrix level, i.e., with better numerical approximation of the 't Hooft limit, estimates of $T_c$ display a decreasing trend; put differently, the temperature range in which the MSS bound is valid expands gradually.
 
The paper is organized as follows. In Sec. \ref{DimenABJM},  we outline and review the various features of the massive deformation of the ABJM model and obtain its reduction from $2+1$ to $0+1$ dimensions by assuming spatially uniform fields. In section \ref{Ansatz1sec}, we introduce our first ansatz configuration, obtain the reduced effective actions, and present the results of the numerical analysis leading to the modeling of the energy dependence of the largest Lyapunov exponent. This is followed by the discussion explaining how we extract the temperature dependence and relate our findings to the MSS conjecture. In Sec. \ref{AnsatzIIsec}, results of an analysis following mainly the same steps of Sec. \ref{Ansatz1sec} are presented for another ansatz configuration. Several details of the calculations are relegated to the Appendixes \ref{AppA} and \ref{AppB}. We conclude in Sec. \ref{con} by briefly summarizing our results and indicating some directions for future studies.

\section{Reduction of mass-deformed ABJM mode to $0+1$ dimensions}
\label{DimenABJM}

We start with writing out the action for the bosonic part of the mass-deformed ABJM model. This is given as \cite{Hosomichi:2008jb, Gomis_2008} 
\begin{multline}
S_{ABJM} = \int d^3 x \frac{k}{4\pi}\epsilon^{\mu\nu\lambda}\Tr(A_{\mu}\partial_{\nu}A_{\lambda}+\frac{2i}{3}A_{\mu}A_{\nu}A_{\lambda}-\hat{A}_{\mu} {\partial_{\nu}}\hat{A}_{\lambda}
-\frac{2i}{3}\hat{A}_{\mu}\hat{A}_{\nu}\hat{A}_{\lambda}) \\
-\Tr|D_\mu Q^\alpha|^2-\Tr|D_\mu R^{\alpha}|^2 - V \,,
\label{SABJM}
\end{multline}
where $A_{\mu}$ and $\hat{A}_{\mu}$ ($\mu:0,1,2$) are two distinct gauge fields transforming under the $SU(N)_k$ and $SU(N)_{-k}$ gauge transformations, respectively. The subscripts $\pm k \in {\mathbb Z}$ label the level of the Chern-Simons terms associated to these gauge fields. The potential term is given as 
\be
V=\Tr(|M^{\alpha}|^2+|N^{\alpha}|^2) \,,
\label{V}
\ee
where 
\beqa
M^\alpha=\mu Q^{\alpha}+\frac{2\pi}{k}(2Q^{[\alpha}Q_{\beta}^{\dagger}Q^{\beta]}+R^{\beta}R_{\beta}^{\dagger}Q^\alpha-Q^\alpha R_\beta^\dagger R^{\beta} +2Q^\beta R_{\beta}^\dagger R^\alpha-2R^\alpha R_{\beta}^\dagger Q^\beta) \,, \nn \\ 
N^\alpha= -\mu R^{\alpha}+\frac{2\pi}{k}(2R^{[\alpha}R_{\beta}^{\dagger}R^{\beta]}+Q^{\beta}Q_{\beta}^{\dagger}R^\alpha-R^\alpha Q_\beta^\dagger Q^{\beta} +2R^\beta Q_{\beta}^\dagger Q^\alpha-2Q^\alpha Q_{\beta}^\dagger R^\beta) \,.
\label{MN}
\eeqa
In this expression  $(Q^\alpha, R^\alpha) := C^I$ with ($\alpha:1,2$) and ($I:1,2,3,4$) are complex bifundamental scalar fields; i.e., they transform as $Q^\alpha \rightarrow U_L Q^\alpha U_R$,  $R^\alpha \rightarrow U_L R^\alpha U_R$, where $(U_L, U_R) \in SU(N)_k \times SU(N)_{-k} $. The covariant derivatives are given as 
\beqa
D_{\mu} Q^\alpha &=& \partial_{\mu}Q^\alpha +i A_\mu Q^\alpha - i Q^\alpha \hat{A}_{\mu} \,, \nn \\
D_{\mu} R^\alpha &=& \partial_{\mu}R^\alpha +i A_\mu R^\alpha - i R^\alpha \hat{A}_{\mu} \,,
\eeqa
and $\mu$ stands for the mass of the fields $(Q^\alpha, R^\alpha)$. $e^{i S_{ABJM}}$ is invariant under the gauge group $SU(N)_k \times SU(N)_{-k}\,,$ provided that $k \in {\mathbb Z}$. The latter is the level quantization of the Chern-Simons couplings in the action. In (\ref{MN}), we use the notation
\be
Q^{[\alpha}Q_{\beta}^{\dagger}Q^{\beta]}=Q^\alpha Q_\beta^\dagger Q^\beta-Q^\beta Q_\beta^\dagger Q^\alpha \,,
\ee
and likewise for $R^\alpha $'s. 

The ABJM model has the global $SU(4)_R \times U(1)_R$ R-symmetry group, which is broken down to $SU(2) \times SU(2)\times U(1)_A \times U(1)_B \times {\mathbb Z}_2$ by the mass deformation terms given in $M^\alpha$ and $N^\alpha$. If there is no mass deformation, it is suitable to formulate the theory in terms of the complex scalar fields $C^I$, which transform under the four-dimensional fundamental representation of the $SU(4)_R$ factor and carry $U(1)_R$ charge $+1$. In the mass-deformed model, $Q^{\alpha}$ transform under the first and $R^\alpha$ transform under the second of the $SU(2)$ factors of the R-symmetry group, and under $U(1)_A\,,$ they have the charges $1\,,-1$, respectively, while under $U(1)_B$, they both have charge $1$, and the ${\mathbb Z}_2$ factor serves to exchange  $Q^{\alpha}$ and $R^{\alpha}$.

To dimensionally reduce $S_{ABJM}$ to $0+1$ dimensions, we declare that all fields are independent of the spatial coordinates and depend on time only. Consequently, all partial derivatives with respect to the spatial coordinates vanish. We may introduce the notation $A_\mu \equiv (A_0,X_i)$, $\hat{A}_\mu \equiv (\hat{A}_0, \hat{X}_i)$ with $(i=1,2)$. Spatial and time components of the covariant derivative are then
\beqa
D_i  Q^{\alpha} &=& iX_i Q^{\alpha}-i Q^{\alpha} \hat{X_i} \,, \quad D_i R^{\alpha} = i X_i R^{\alpha} -i R^{\alpha} \hat{X_i} \,,  \nn \\
D_0 Q^{\alpha} &=& \partial_{0} Q^{\alpha} +i A_0 Q^{\alpha} -i Q^{\alpha} \hat{A_{0}} \,, \quad D_0 R^{\alpha} = \partial_{0} R^{\alpha} +i A_0 R^{\alpha} -i R^{\alpha} \hat{A_{0}} \,, 
\eeqa
and the action takes the following form:
\begin{multline}
S_{ABJM-R} = N \int d t \,\,  \frac{k}{4\pi}\Tr(-\epsilon^{ij} X_i \dot{X}_j + i \epsilon^{ij} A_0 \lbrack X_i \,,X_j \rbrack) - \frac{k}{4\pi}\Tr(-\epsilon^{ij}\hat{X}_i\dot{\hat{X}}_j + i \epsilon^{ij} \hat{A_0} \lbrack \hat{X_i} \,, \hat{X_j}\rbrack)  \\
+ \Tr(|D_0  Q^\alpha|^2) -\Tr(|D_i Q^\alpha|^2) + \Tr(|D_0  R^\alpha|^2) -\Tr(|D_i R^\alpha|^2) - V \,.
\label{ABJMR1}
\end{multline}
Expressing the Chern-Simons parts of this action in terms of the covariant derivatives ${\mathcal D}_0 {X}_i := \partial_0 X_i - i \lbrack A_0 \,, X_i \rbrack$, and ${\hat {\mathcal D}}_0 {\hat{X}}_i := \partial_0 \hat{X}_i - i \lbrack {\hat A}_0 \,, \hat{X}_i \rbrack$, we may as well write
\begin{multline}
S_{ABJM-R} = N \int d t \, \, - \frac{k}{4\pi} \Tr(\epsilon^{ij} X_i {\mathcal D}_0 {X}_j) + \frac{k}{4\pi}\Tr(\epsilon^{ij} \hat{X}_i  {\hat {\mathcal D}}_0 {\hat{X}}_j ) + \Tr(|D_0  Q^\alpha|^2) 
\\
-  \Tr(|D_i Q^\alpha|^2)  +  \Tr(|D_0  R^\alpha|^2) -\Tr(|D_i R^\alpha|^2) - V \,.
\label{ABJMR2}
\end{multline}
In (\ref{ABJMR1}) and (\ref{ABJMR2}),  it is understood that all fields depend on time only. We have readily written the reduced action in the 't Hooft limit. The latter is defined as follows. While reducing from $2+1$ to $0+1$ dimensions, we have integrated over the two-dimensional space whose volume may be denoted, say, by $V_2$. Therefore, we may introduce\footnote{Note that in the original model, i.e., in $2+1$ dimensions, the 't Hooft coupling is identified as $\lambda_{'t \, Hooft} = \frac{N}{k}$ held fixed with $N, k \rightarrow \infty$ \cite{Nastase:2015wjb}. In the reduced model, too, we may define $\tilde{\lambda}_{' t \, Hooft} := \frac{N}{ k V_2} = \frac{\lambda_{' t \, Hooft}}{k} $ held fixed with $N, V_2 \rightarrow \infty$, while, in contrast, $k$ can remain finite. In this case, scaling $\tilde{\lambda}_{' t \, Hooft}$ to $\frac{1}{k}$  is the same as scaling $\lambda_{' t \, Hooft}$ to unity.}  $\lambda_{' t \, Hooft} := \frac{N}{V_2}$ and require that it remains finite in the limit $V_2 \rightarrow \infty $ and $N \rightarrow \infty$. In  $S_{ABJM-R}$, we have scaled $\lambda_{'t \, Hooft}$ to unity. If needed, it is possible to restore $\lambda_{'t \, Hooft}$ back in $S_{ABJM-R}$ by performing the scalings $X_i \rightarrow \lambda^{-1/2}X_i$, $\hat{X_i} \rightarrow \lambda^{-1/2} \hat{X_i}$, $A_0 \rightarrow \lambda^{-1/2}A_0$, $\hat{A_0} \rightarrow \lambda^{-1/2}\hat{A_0}$, $Q_\alpha \rightarrow \lambda^{-1/4}Q_\alpha$, $R_\alpha \rightarrow \lambda^{-1/4}R_\alpha$, $\mu \rightarrow \lambda^{-1/2}\mu\,,$ and $t \rightarrow \lambda^{1/2} t $ . Let us note also that $S_{ABJM-R}$ is manifestly gauge invariant under the $SU(N)_k \times SU(N)_{-k}$ gauge symmetry.\footnote{In particular, let us note that pure CS action is indeed manifestly gauge invariant in $0+1$ dimensions as opposed to the non-Abelian CS action in $2+1$ dimensions, which is not.  The latter gives rise to the level quantization of the CS coupling, i.e., $k \in {\mathbb Z} $. In fact, after the reduction of the CS terms to $0+1$ dimensions but prior to introducing the 't Hooft parameter $\lambda_{'t \, Hooft}$, the effective CS coupling is simply $\kappa := \frac{1}{4 \pi} k V_2$ and is no longer an integral multiple of $\frac{1}{4 \pi}$ due to the arbitrary volume $V_2$ of the two-dimensional compact space we have integrated over. This is consistent with the fact that CS term in $0+1$ dimensions is gauge invariant and therefore its coupling is not level quantized. The latter also follows from the fact that $\pi_{1}(SU(N)) = 0$ and the general considerations on the gauge symmetry properties of $e^{i S_{CS}}$ which may be found, for instance, in Refs. \cite{Bal, Dunne}.} 

The ground states of this reduced model are the same as that of the original model and given by configurations minimizing the potential $V$ in \eqref{V}. Since the latter is positive definite, its minimum is zero and is given by the configuration
\be
M^\alpha=0=N^\alpha \,.
\label{V0s}
\ee
There are two immediate solutions to \eqref{V0s}, which are given as
\beqa
R^\alpha=&c \, G^\alpha\,, \quad  Q^\alpha =0 \,, \nn \\
R^\alpha=&0 \,, \quad Q^\alpha=c \, G^\alpha \,,
\label{antz2}
\eeqa
where $G^\alpha$ are the GRVV matrices \cite{Gomis_2008,Nastase:2015wjb} defining a fuzzy 2-sphere \cite{Balachandran:2005ew} at the matrix level $N$ and $c = \sqrt{\frac{k\mu}{4\pi}}$. Let us note in passing that $c=0$ gives a trivial solution in which both the fields $Q^\alpha$ and $R^\alpha$ vanish and is of no interest to us in what follows. Explicitly, $G^\alpha$ are given as \cite{Gomis_2008}

\begin{align}\label{GRVValb}
	\begin{split}
		(G^1)_{mn}&=\sqrt{m-1}\,\delta_{m,n} \,,\\
		(G^2)_{mn}&=\sqrt{N-m}\,\delta_{m+1,n} \,, \\
		(G_1^\dagger)_{mn}&=\sqrt{m-1}\,\delta_{m,n} \,, \\
		(G_2^\dagger)_{mn}&=\sqrt{N-n}\,\delta_{n+1,m} \,,
	\end{split}
\end{align}
with $m,n = 1 \,, \cdots \,, N$, and they fulfill the relation
\be
G^{\alpha} = G^\alpha G_\beta^\dagger G^\beta-G^\beta G_\beta^\dagger G^\alpha \,.
\label{GRVV}
\ee

We may notice at this stage that it is possible to work in the gauge with $A_0=0$ and $\hat{A}_0=0$.  Evaluating the variations of $S_{ABJM-R}$ with respect to $A_0$ and $\hat{A}_0$, we find the Gauss-law constraint is given  by the two equations 
\begin{align}\label{gausslaw}
	\begin{split}
		\frac{k}{2\pi}[X_1,X_2] + \dot{Q}^\alpha Q_\alpha^{\dagger} -Q^\alpha \dot{Q}_\alpha^{\dagger} + \dot{R}^\alpha R_\alpha^{\dagger} -R^\alpha \dot{R}_\alpha^{\dagger}&= 0 \,, \\
		-\frac{k}{2\pi}[\hat{X}_1,\hat{X}_2] - Q_\alpha^{\dagger} \dot{Q}^\alpha +\dot{Q}_\alpha^{\dagger}Q^\alpha  - R_\alpha^{\dagger} \dot{R}^\alpha +\dot{R}_\alpha^{\dagger} R^\alpha&= 0 \,.
	\end{split}
\end{align}

It is also useful to note that the Hamiltonian takes the form 
\be
H = \Tr \left( \frac{1}{N} |P_Q^\alpha|^2 +  \frac{1}{N}  |P_R^\alpha|^2 +N  |D_i Q^{\alpha}|^2 + N | D_i R^{\alpha}|^2 \right ) + N V  \,,
\ee
where 
\be
P_Q^\alpha =\frac{\partial L}{\partial {\dot Q}^\alpha} = N {\dot Q}^{\alpha \dagger} \,, \quad  P_R^\alpha =\frac{\partial L}{\partial {\dot R}^\alpha}= N {\dot R}^{\alpha \dagger} \
\ee
are the conjugate momenta associated to $Q_\alpha$ and $R_\alpha$, respectively. It is straightforward to see that the Hamiltonian for the CS part of the action vanishes identically as expected \cite{Dunne}.

Let us consider the scaling transformation
\be 
(Q_\alpha \,, R_\alpha) \rightarrow  (\rho^{-1/2} \, Q_\alpha \,, \rho^{-1/2} \, R_\alpha) \,, \quad (X_i \,, \hat{X}_i) \rightarrow  (\rho^{-1} \, X_i \,, \rho^{-1}\, \hat{X}_i) \,, \quad t \rightarrow \rho \, t \,, 
\ee
where $\rho$ is an arbitrary positive constant. Under this transformation, we have $(P_Q^\alpha\,, P_R^\alpha) \rightarrow (\rho^{-3/2} \, P_Q^\alpha \,, \rho^{-3/2} \, P_R^\alpha)$ and $V |_{\mu = 0} \rightarrow \rho^{-3} \, V |_{\mu =0}\,$. Therefore, the energy scales as $E \rightarrow \rho^{-3} E $. Since the Lyapunov exponent has the dimensions of inverse time, we see that it scales as 
\be
\lambda_L \propto E^{1/3} 
\ee
in the massless limit. In the ensuing sections, we will see that this scaling of the Lyapunov exponents with energy is essentially preserved after taking the mass deformations into account.  

We are now in a position to propose ansatz configurations, through which we will be able to explore the emerging chaotic dynamics. We will consider two different ansatz configurations involving the GRVV matrices and satisfying the Gauss-law constraints. Both of these ansatz configurations involve collective time dependence and are introduced via real functions of time. 

\section{Ansatz I and the effective action}

\label{Ansatz1sec}

The first matrix configuration we focus on is specified as
\begin{align}
	\begin{split}\label{ansatz11}
		X_i &= \alpha(t)\text{diag}((A_i)_1,(A_i)_2,...,(A_i)_N)  \,, \\
		\hat{X}_i &= \beta(t)\text{diag}((B_i)_1,(B_i)_2,...,(B_i)_N) \,, \\
		Q_\alpha & = \phi_\alpha(t)G_\alpha, \quad R_\alpha=0 \,, 
	\end{split}
\end{align}
where $(A_i)_m$, $(B_i)_m$ are constants and $i =1,2$, $m = 1,2,...,N$ and $\alpha =1,2$. Thus, $X_i$ and $\hat{X_i}$ are taken as diagonal matrices. No sum over  the repeated index $\alpha$ is implied in the last line of \eqref{ansatz11}. Here $\phi_\alpha(t)$, $\alpha(t)$, $\beta(t)$ are real functions of time, and the Gauss-law constraint given in the Eq. \eqref{gausslaw} is easily seen to be satisfied by this choice of the matrices. 

Evaluating the equations of motion for $\alpha(t)$ and $\beta(t)$, we find that the emerging coupled equations have only one possible real solution and that is the trivial solution given simply as $\alpha(t)= \beta(t)=0$. This result is proved in Appendix (\ref{AppA}). Henceforth, setting $X_i$ and $\hat{X_i}$ to zero, inserting last line of \eqref{ansatz11} in the action \eqref{ABJMR2}, and performing the traces over the GRVV matrices at the level of $N \times N$ matrices, we obtain the reduced Lagrangian
\begin{align}
	\begin{split}
		L_N =& N^2(N-1)\Big(\frac{1}{2}{\dot \phi}_1^2+\frac{1}{2} {\dot \phi}_2^2-\frac{1}{2}\mu^2(\phi_1^2+\phi_2^2)-\frac{8\pi\mu}{k}\phi_1^2\phi_2^2-\frac{8\pi^2}{k^2}\phi_1^4\phi_2^2-\frac{8\pi^2}{k^2}\phi_1^2\phi_2^4 \Big) \,.
	\end{split}
\end{align}
The corresponding Hamiltonian is
\begin{align}\label{hamiltonian1}
\begin{split}
		H_N(\phi_1,\phi_2\,, & p_{\phi_1},p_{\phi_2}) = \frac{p_{\phi_1}^2}{2N^2(N-1)} + 
		\frac{p_{\phi_2}^2}{2N^2(N-1)}\\
		&+N^2(N-1)\left(\frac{1}{2}\mu^2(\phi_1^2+\phi_2^2)+\frac{8\pi\mu}{k}\phi_1^2\phi_2^2+\frac{8\pi^2}{k^2}\phi_1^4\phi_2^2+\frac{8\pi^2}{k^2}\phi_2^4\phi_1^2\right)\\
		&\qquad\hspace{0.5em} \quad=:\frac{p_{\phi_1}^2}{2N^2(N-1)}+\frac{p_{\phi_2}^2}{2N^2(N-1)}+V_N(\phi_1,\phi_2) \,.
\end{split}
\end{align}
where $V_N(\phi_1,\phi_2)$ introduced in the second line denotes the potential of this reduced system and defined by the relevant expression in the first line. For $k>0$, we see that $V_N(\phi_1,\phi_2)$ is clearly positive definite, while for $k<0$, this is not manifest, but it is indeed so since $V_N(\phi_1,\phi_2)$ is obtained from $V$ in \ref{V}. Hence the minimum of $V_N(\phi_1,\phi_2)$ is zero in both cases.

Let us note that, in the $\mu \rightarrow 0$ limit, we have $H_N \rightarrow \rho^{-3} H_N$ under the scaling $(\phi_1 \,, \phi_2) \rightarrow (\rho^{-1/2} \, \phi_1 \,, \rho^{-1/2} \, \phi_2)$ and $t \rightarrow \rho \, t $, as can be readily expected in view of the discussion given at the end of the previous section.

To explore the dynamics of the model, we calculate the Hamiltonian equations of motion. These take the form
\begin{subequations}\label{e.o.m}
	\begin{align}
		&\dot{\phi}_1-\frac{p_{\phi_1}}{N^2(N-1)}=0\,,\\
		&\dot{\phi}_2-\frac{p_{\phi_2}}{N^2(N-1)}=0\,,\\
		&\dot{p}_{\phi_1}+N^2(N-1)\left(\mu^2 \phi_1+\frac{16\pi \mu}{k}  \phi_1\phi_2^2 + \frac{16\pi^2}{k^2} \phi_1\phi_2^4 +\frac{32\pi^2}{k^2}\phi_1^3\phi_2^2 \right)=0\,,\\
		&\dot{p}_{\phi_2}+N^2(N-1)\left(\mu^2{\phi}_2+\frac{16\pi \mu}{k} \phi_1^2{\phi}_2 + \frac{16\pi^2}{k^2} \phi_1^4{\phi}_2 +\frac{32\pi^2}{k^2} \phi_1^2{\phi}_2^3 \right)=0.
	\end{align}
\end{subequations}

In what follows, we will explore the dynamics emerging from the equations at $\mu =1$ at several different matrix levels $N$ and the CS coupling $k$.

To gain some immediate information on the system, it is useful to explore its fixed points and also investigate the stability around these points at the linear level. Details of this analysis are provided in Appendix \ref{AppB}. In brief, for $k > 0$, the only fixed point of this Hamiltonian system is given as $(\phi_1,\phi_2,p_{\phi_1},p_{\phi_2})\equiv(0,0,0,0)$ with a vanishing fixed-point energy. Analysis in Appendix \ref{AppB} shows that this fixed point is of borderline type, meaning that the linear level analysis is inconclusive to identify it as either stable or unstable character, and we do not attempt to perform a higher-order analysis. For $k < 0$, we find that there are several fixed points, some of which are still of borderline type. Nevertheless, the set of fixed points given as $ \left(\pm (\mp)\frac{\sqrt{-k}}{2\sqrt{3\pi}},\pm\frac{\sqrt{-k}}{2\sqrt{3\pi}},0,0\right)$ is of unstable type with energies $E_F = N^2 (N-1) \frac{5 |k  \mu^3 |}{108 \pi}$ (for $k  \mu < 0 $), as calculated in Appendix B. This may be taken as the first indication to expect the dynamics of $H_N$ to be chaotic, since the latter is usually associated to the presence of unstable fixed points in the phase space \cite{Ott, Hilborn, Campbell, Percival}. In fact, the systems do not tend to exhibit any appreciable chaos at energies below that of the unstable fixed points, and depending on the structure of the potential, even at energies exceeding the latter, phase space may have comparable number of quasiperiodic and chaotic trajectories; i.e., at low energies, chaos and quasiperiodic motion can coexist. In particular, a randomly picked initial condition  may correspond to either a quasiperiodic or a chaotic trajectory. Therefore, it is important to pay attention to this fact in the computation of the Lyapunov exponents, and we will do so in the ensuing sections. 

\subsection{Chaotic dynamics and the Lyapunov exponents}
\label{lyap1}

Lyapunov exponents are useful to determine the sensitivity of a system to given initial conditions. More precisely, they measure the exponential growth in perturbations and therefore give a reliable way to establish the presence of chaos in a dynamical system \cite{Ott,Hilborn,Campbell, Percival}. For a Hamiltonian system, if we denote the perturbations in the phase-space coordinates ${\bm g}(t) \equiv (g_1(t), g_2(t)\,,\cdots\,, g_{2N}(t))$ by $\delta {\bm g}(t)$, then we may conclude that the system is chaotic if, at large $t$, $\delta {\bm g}(t)$ deviates exponentially from its initial value at $t= t_0$: $||\delta {\bm g}(t)|| = e^{\lambda (t-t_0)} ||\delta {\bm g}(t_0)||$. Here, $\lambda $ are called the Lyapunov exponents, and there are $2n$ of them for a phase space of dimension $2n$. Let us also note that this description is in parallel with the statement that even slightly different initial conditions give trajectories in the phase space, which are exponentially diverging from each other and hence lead to chaos. In a dynamical system, the presence of at least one positive Lyapunov exponent is sufficient to conclude the presence of chaotic motion. In Hamiltonian systems, due to the symplectic structure of the phase space, Lyapunov exponents appear in $\lambda_i$ and $-\lambda_i$ pairs, and a pair of the Lyapunov exponents vanishes, as there is no exponential growth in perturbations along the direction of the trajectory specified by the initial condition and the sum of all the Lyapunov exponents is zero as a consequence of Liouville's theorem. These facts are well known, and their details may be found in many of the excellent books on chaos \cite{Ott,Hilborn,Campbell, Percival}.  The phase space for the Hamitonians $H_N$ considered in this paper are all four dimensional. From the general considerations summarized above, it is clear that the emerging chaotic dynamics of these models are governed by the largest (and only) positive Lyapunov exponent at given values of the parameters $k$, $\mu$, and $N$. 

To give a certain effectiveness to the random initial condition selection process, we adapt and use the simple approach we have developed in Ref. \cite{Baskan:2019qsb}. We briefly explain this next. Let us denote a generic set of initial conditions at $t=0$ by $(\phi_1(0),\phi_2(0),p_{\phi_1}(0),p_{\phi_2}(0))$. First of all, we generate four random numbers and denote three of them as $\omega_i$ $(i=1,2,3)$ and define $\Omega_i = \frac{\omega_i}{\sqrt{\omega_i^2}}\sqrt{E}$, where $E$ is the energy of the system. We denote the last random number as $\omega_4$. Clearly, we have $\sum_{i=1}^{3}\Omega^2_i=E$. With the help of this relation and the energy functional, i.e., Hamiltonian given in \eqref{hamiltonian1}, initial conditions are picked randomly in the form 
\begin{align}
p_{\phi_1}(0) =  \pm N \sqrt{ 2 (N-1)} \Omega_1 \,, \quad p_{\phi_2}(0) = \pm N \sqrt{2(N-1)} \Omega_2 \,, \quad V_N (\phi_1(0) \,, \phi_2(0)) = \Omega_3^2 \,.
\end{align}
Finally, we select either $\phi_1(0)$ or $\phi_2(0)$ as $\omega_4$. Since $V_N$ is invariant under $\phi_1 \leftrightarrow \phi_2$ exchange, which one of the two we select is immaterial. In our calculations, we take $\phi_2(0)=\omega_4$; then, $\phi_1(0)$ is given by the solution of
\begin{align}\label{initials3}
V_N\left(\phi_1(0),\omega_4\right)-\Omega^2_3=0 \,.
\end{align}
For some randomly picked values of $\omega_4$, Eq. \eqref{initials3} may not have a real solution. However, the code runs and randomly picks another $\omega_4$ until a real solution for $\phi_1(0)$ is obtained. 

Similar to the analysis performed for the Yang-Mills matrix models with massive deformations presented in Ref. \cite{Asano:2015eha},  we set up and run a MATLAB code, which numerically solves the Hamilton equations of motion given in \eqref{e.o.m} at different matrix levels. We run this code $40$ times for $k  \geq 1$  and $100$ times for $ k \leq-1 $  with randomly selected initial conditions at a given energy value $E$ and matrix level $N$ and calculate the average for each and every Lyapunov exponent from all runs a the final time. In the simulation, we take a time step of $0.25$ and run the code from time $0$ to $3000$. Our code checks if the largest Lyapunov exponent has a value below a certain threshold at $t = 3000$ and does not include it in the averaging over the initial conditions. In our computations, we picked this threshold as $0.05$ after a number of numerical trials.\footnote{Except for the case $k=-2$, $N=10$, for which we picked the threshold as $0.1$.} Let us note that Lyapunov exponents below this threshold at large time ($t = 3000$ in our simulations) correspond essentially to the quasiperiodic trajectories in the phase space, which do not exhibit chaos but may have comparatively small or large periods and therefore usually have very small but nonvanishing Lyapunov exponents at large time,\footnote{The number of such trajectories is very few for the configuration due to ansatz I and  essentially becomes zero with increasing energy, while for ansatz II, roughly $\approx 1/5$ to $\approx 1/10$ of the initial conditions lead to quasiperiodic orbits at low energies, but their number also becomes zero with increasing energy.} and in the manner just described, we exclude them in order to obtain more precise values for the largest Lyapunov exponents of the chaotic trajectories in the phase space. In particular, we focus on $H_N$ for $N = 5,10,15,20,25$ at several different values of the energy. 

\subsection{Dependence of the largest Lyapunov exponent on energy}

Since we are working in the 't Hooft limit, it is useful to consider the dependence of the largest Lyapunov exponent, $\lambda_L$, on $E/N^2$ rather than on $E$, to capture the main features of the chaotic dynamics emerging from the family of Hamiltonians $H_N$ and subsequently relate it to the temperature of these systems via the use of virial and equipartition theorems.

\vskip 1em

\textit{Case i : $k\geq1$}

\vskip 1em

\begin{figure}[!htb]
	\begin{subfigure}[!htb]{.5\textwidth}
		\centering
		\includegraphics[width=6.6cm]{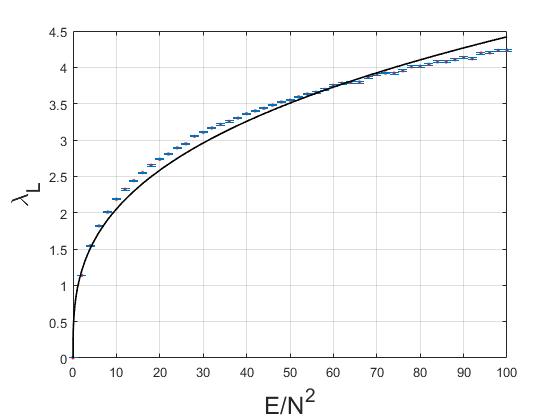}
		\caption{$N=5$}
		\label{fig:len5}
	\end{subfigure}
	\begin{subfigure}[!htb]{.5\textwidth}
		\centering
		\includegraphics[width=6.6cm]{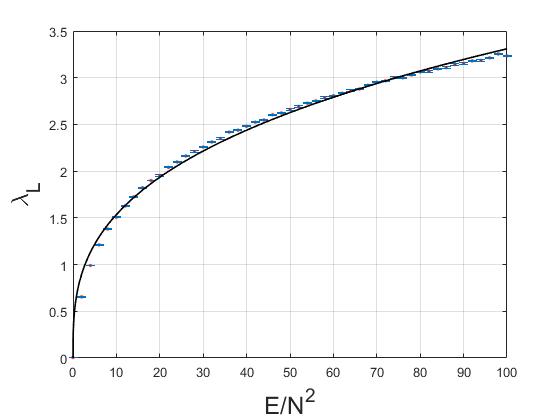}
		\caption{$N=10$}
		\label{fig:len101}
	\end{subfigure}
	\begin{subfigure}[!htb]{.5\textwidth}
		\centering
		\includegraphics[width=6.6cm]{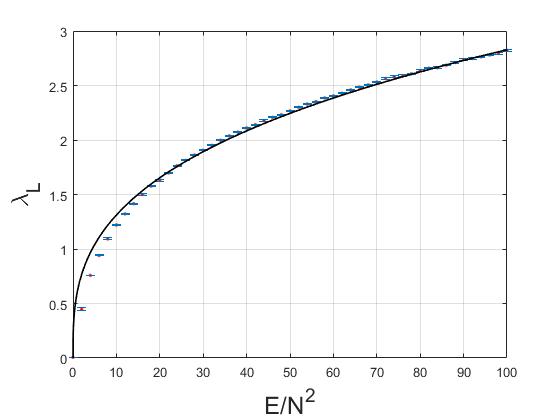}
		\caption{$N=15$}
		\label{fig:len151}
	\end{subfigure}
	\begin{subfigure}[!htb]{.5\textwidth}
		\centering
		\includegraphics[width=6.6cm]{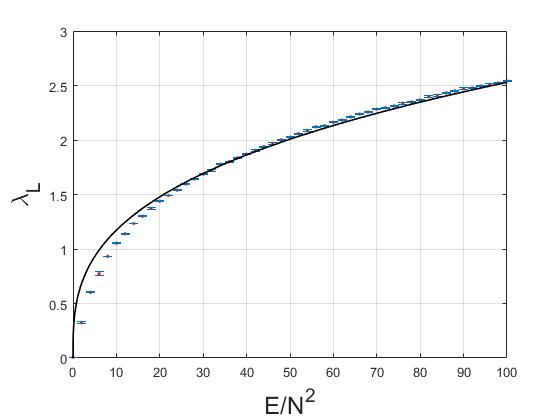}
		\caption{$N=20$}
		\label{fig:len201}
	\end{subfigure}
	\begin{center}
		\begin{subfigure}[!htb]{.5\textwidth}
			\centering
			\includegraphics[width=6.6cm]{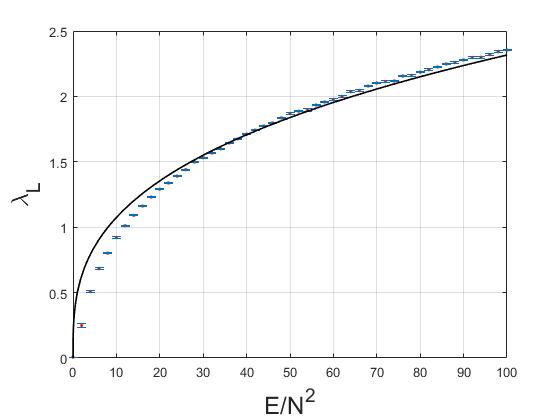}
			\caption{$N=25$}
			\label{fig:len25}
		\end{subfigure}
	\end{center}
	\caption{Largest Lyapunov exponent and the best-fitting curves in the form $\lambda_L=\alpha_N(\frac{E}{N^2})^{1/3}$ at $k=1$.}
	\label{fig1}
\end{figure}

\begin{figure}[!htb]
	\begin{subfigure}[!htb]{.5\textwidth}
		\centering
		\includegraphics[width=6.6cm]{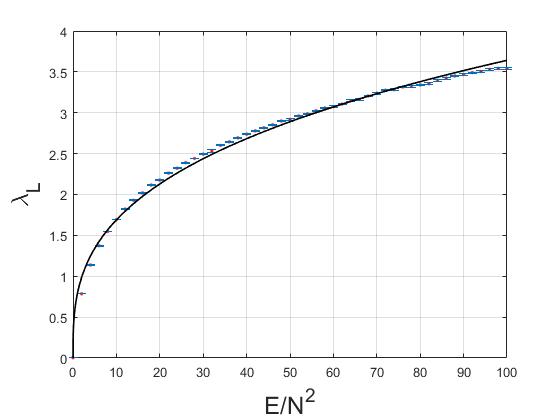}
		\caption{$N=5$}
		\label{fig:len5k2}
	\end{subfigure}
	\begin{subfigure}[!htb]{.5\textwidth}
		\centering
		\includegraphics[width=6.6cm]{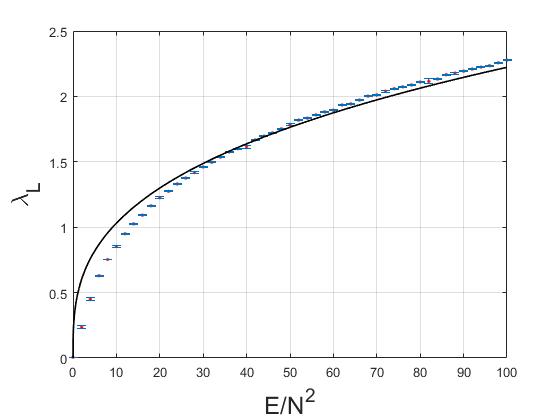}
		\caption{$N=15$}
		\label{fig:len15k2}
	\end{subfigure}
	\begin{center}
		\begin{subfigure}[!htb]{.5\textwidth}
			\centering
			\includegraphics[width=6.6cm]{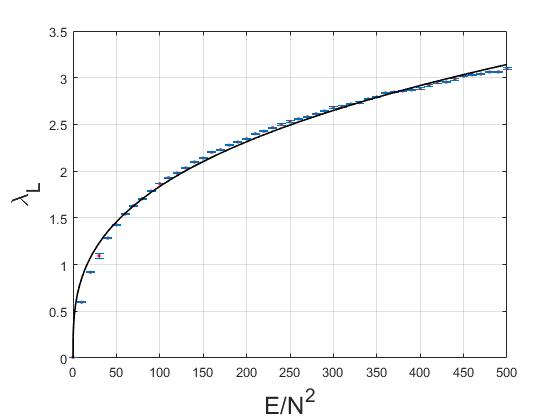}
			\caption{$N=25$}
			\label{fig:len25k2}
		\end{subfigure}
	\end{center}
	\caption{Largest Lyapunov exponent and the best-fitting curves in the form $\lambda_L=\alpha_N(\frac{E}{N^2})^{1/3}$ at $k=2$.}
	\label{figk2}
\end{figure}

In this case, to capture the $\lambda_L \propto E^{1/3}$ dependence of the Lyapunov exponent anticipated by the scaling argument given in Sec. $2$, we find that it is sufficient to choose $E/N^2$ in the interval $(0, 100)$. Lyapunov exponent data and the best-fitting curves of the form $\lambda_L=\alpha_N(\frac{E}{N^2})^{1/3}$ are given in Fig. \eqref{fig1} for $k=1$ and $N=5,10,15,20,25$. For $k=2$, at sufficiently low matrix levels, the $E/N^2$ interval can still be taken as $(0, 100)$, while for $N > 20$, it turns out to be better to stretch it to a wider range, and in Fig. \ref{fig:len25k2}, we take it to be $(0, 500)$. Lyapunov exponent data and the best-fitting curves of the form $\lambda_L=\alpha_N(\frac{E}{N^2})^{1/3}$ are given in the Fig. \eqref{figk2}.

\begin{figure}[!htb]
	\begin{subfigure}[!htb]{.5\textwidth}
		\centering
		\includegraphics[width=6.6cm]{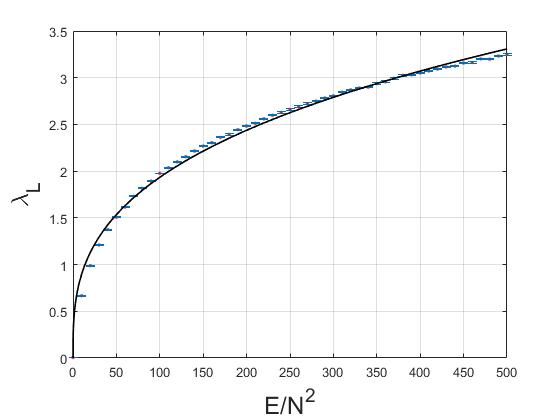}
		\caption{$k=5$}
		\label{fig:n10k5}
	\end{subfigure}
	\begin{subfigure}[!htb]{.5\textwidth}
			\centering
				\includegraphics[width=6.6cm]{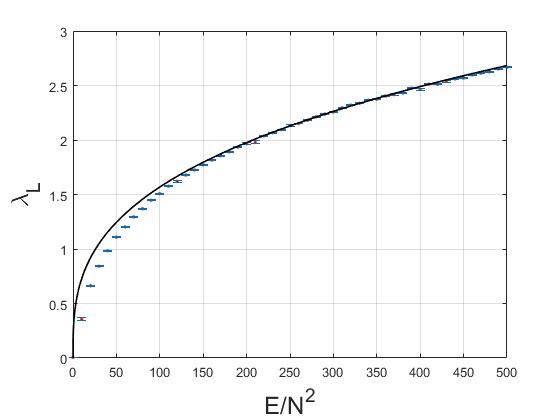}
				\caption{$k=10$}
				\label{fig:n10k10}
			\end{subfigure}
	\caption{Largest Lyapunov exponent and the best-fitting curves in the form $\lambda_L = \alpha_N(\frac{E}{N^2})^{1/3}$ for $N=10$ at $k=5,10$.}
	\label{fign10}
\end{figure}

If we further increase the $k$ value, we also need to inspect the dependence of $\lambda_L$  to $E/N^2$ in a sufficiently large range of the latter.  For instance, in Fig. \eqref{fign10}, we depict the Lyapunov data for $0 \leq E/N^2 \leq 500$, at the matrix level $N=10$ and for $k=5, 10$. At higher matrix levels $N$ and/or larger values of $k$, it is necessary to further increase the range of $E/N^2$ in order to clearly observe the $E^{1/3}$ dependence of $\lambda_L$ via the best-fitting curves in the form $\lambda_L=\alpha_N(\frac{E}{N^2})^{1/3}$. Coefficients $\alpha_N$ for the fitting curves in Figs. \eqref{fig1}, \eqref{figk2}, and \eqref{fign10} are provided in Tables \ref{table:tablek1}, \ref{table:tablek2}, and \ref{table:tablen10} given in the next subsection. 
 
\vskip 1em

\textit{Case ii : $k\leq -1$}

\vskip 1em

In this case, we seek best-fitting curves of the form $\lambda_L = \alpha_N (\frac{E}{N^2}- \gamma_N)^{1/3}$ to the Lyapunov data. Motivations for considering this function of $E/N^2$ are twofold. For one, as noted earlier, the energies of the unstable fixed points are nonvanishing in this case and given by $E_F = N^2 (N-1) \frac{5 |k \mu^3|}{108 \pi}$. Since no significant chaos is present for $E \leq E_F$, we expect $\lambda_L$'s to vanish at energies below $E_F$ (indeed, all our numerical computations show that $\lambda_L$ are vanishingly small for $E \leq E_F$). This suggests then that $\gamma_N^{(1)} := \frac{E_F}{N^2}$ and it is determined in terms of $N$, $k$, and $\mu$. In Sec. \ref{Tdependence}, we will see that  the application of the virial theorem to this family of systems motivates the same form for the dependence of $\lambda_L$ on $E/N^2$ with $\gamma_N^{(2)} :=  (N-1) \frac{2}{27 \pi} |k \mu^3|$ via \eqref{minv1} and \eqref{Ineqansatz1}. The numerical values of $\gamma_N^{(1)}$ and $\gamma_N^{(2)}$ are comparable, and for the evaluation of the coefficients $\alpha_N$ of the fitting curves, we use the latter as they tend to produce slightly better fits.    

\begin{figure}[!htb]
	\begin{subfigure}{.5\textwidth}
		\centering
		\includegraphics[width=6.6cm]{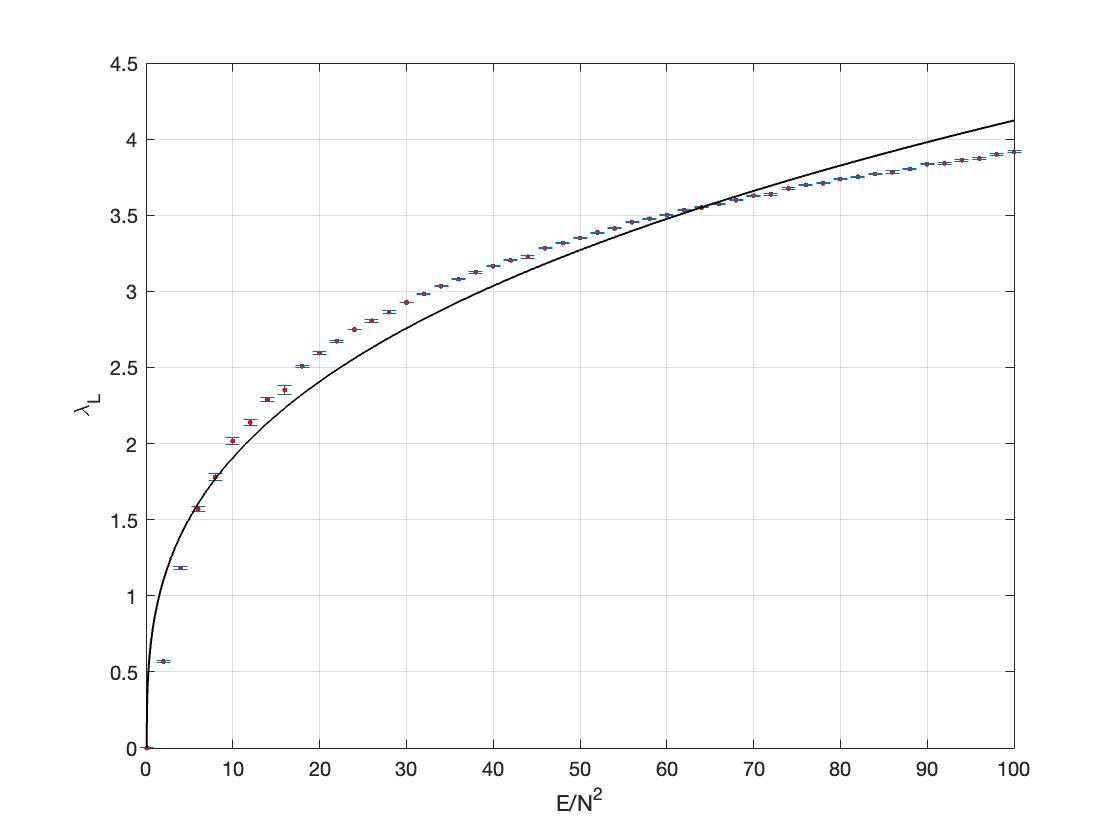}
		\caption{$N=5$}
		\label{fig:len5k-1}
	\end{subfigure}
	\begin{subfigure}{.5\textwidth}
		\centering
		\includegraphics[width=6.6cm]{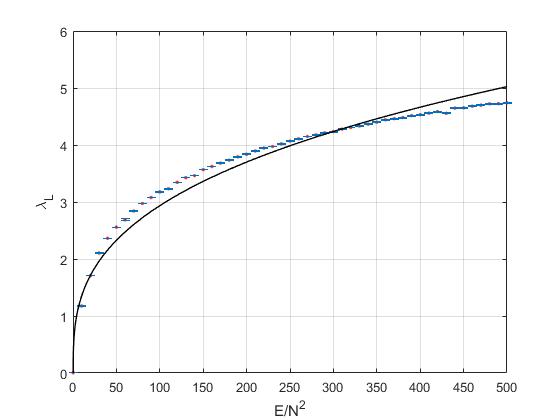}
		\caption{$N=10$}
		\label{fig:len10k-1}
	\end{subfigure}
	\begin{subfigure}{.5\textwidth}
		\centering
		\includegraphics[width=6.6cm]{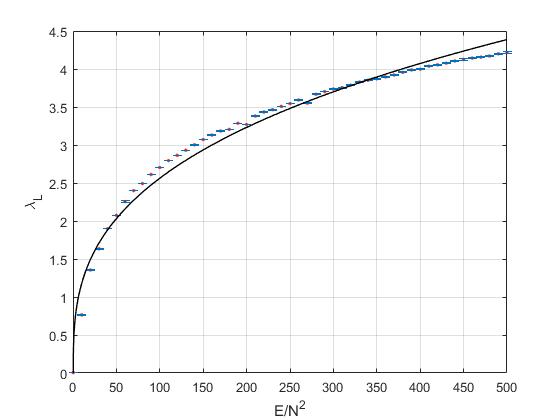}
		\caption{$N=15$}
		\label{fig:len15k-1}
	\end{subfigure}
	\begin{subfigure}{.5\textwidth}
		\centering
		\includegraphics[width=6.6cm]{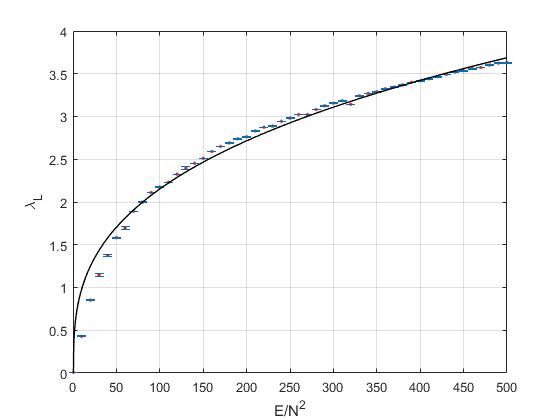}
		\caption{$N=20$}
		\label{fig:len20k-1}
	\end{subfigure}
	\begin{center}
		\begin{subfigure}{.5\textwidth}
			\centering
			\includegraphics[width=6.6cm]{A1k-1N25}
			\caption{$N=25$}
			\label{fig:len25k-1}
		\end{subfigure}
	\end{center}
	\caption{Largest Lyapunov exponent and the best-fitting curves in the form $\lambda_L=\alpha_N(\frac{E}{N^2}-\gamma_N)^{1/3}$ at $k=-1$.}
	\label{fig-1}
\end{figure}

\begin{figure}[!htb]
	\begin{subfigure}[!htb]{.5\textwidth}
		\centering
		\includegraphics[width=6.6cm]{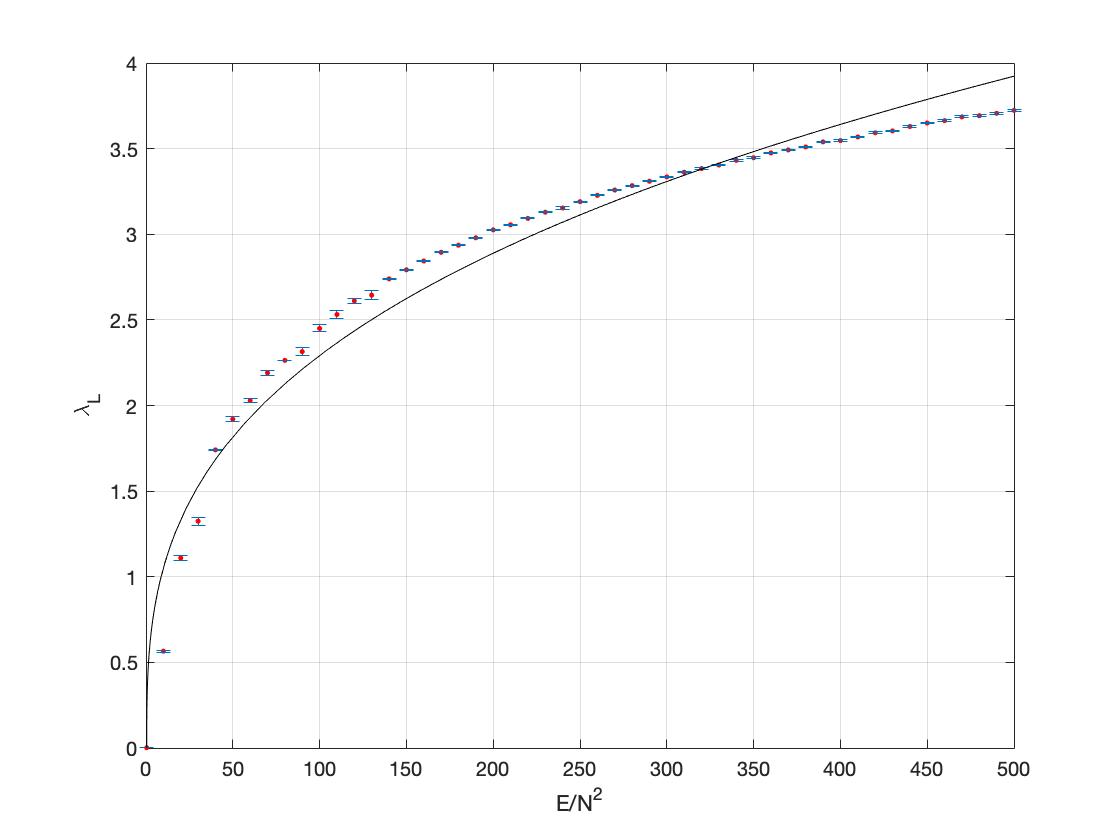}
		\caption{$N=10$}
		\label{fig:len5k-2}
   	\end{subfigure}
	\begin{subfigure}[!htb]{.5\textwidth}
		\centering
		\includegraphics[width=6.6cm]{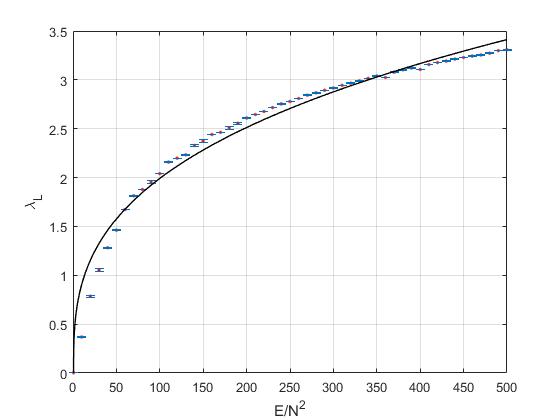}
		\caption{$N=15$}
		\label{fig:len15k-2}
	\end{subfigure}
	\begin{center}
		\begin{subfigure}[!htb]{.5\textwidth}
			\centering
			\includegraphics[width=6.6cm]{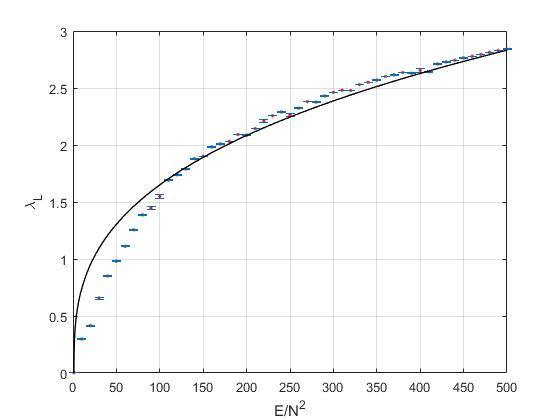}
			\caption{$N=25$}
			\label{fig:len25k-2}
		\end{subfigure}
	\end{center}
	\caption{Largest Lyapunov exponent and the best-fitting curves in the form $\lambda_L=\alpha_N(\frac{E}{N^2}-\gamma_N)^{1/3}$ at $k=-2$.}
	\label{figk-2}
\end{figure}

Both for $k=-1$ and $k =- 2$, we find that the $(0, 500)$ interval for $E/N^2$ is sufficiently well suited to capture the energy dependence of $\lambda_L$.  This is corroborated by the best-fitting curves of the form $\lambda_L=\alpha_N(\frac{E}{N^2}-\gamma_N)^{1/3}$ given in Figs. \eqref{fig-1} and \eqref{figk-2}. Coefficients $\alpha_N$ for the fitting curves in Figs. \eqref{fig-1} \eqref{figk-2} and the respective values of $\gamma_N$ are provided Tables
\ref{table:tablek-1} and \ref{table:tablek-2} in the next subsection.

\subsection{Temperature dependence of the Lyapunov exponent} \label{Tdependence}

In Ref. \cite{Gur-Ari:2015rcq}, temperature dependence of the Largest Lyapunov exponent of the BFSS matrix model in the 't Hooft limit was determined using dimensional analysis to be of the form $\lambda_L \propto (\lambda_{'t\, Hooft}T)^{1/4}$ since $\lambda_{'t\, Hooft}$ and the temperature are the only dimensionful parameters of the model. Let us note that this result is consistent with the fact that for the BFSS model the potential is purely quartic and the system has a scaling symmetry implying that $\lambda_L \propto E^{1/4}$ and hence $\lambda_L \propto T^{1/4}$ temperature dependence by evoking the equipartition theorem. In Ref. \cite{Baskan:2019qsb}, we focused on a mass-deformed Yang-Mills matrix theory, with the same matrix content as the bosonic part of the BFSS model, and a similar analysis is considered where the effects of the mass deformations were taken into account in a simple way. For the present matrix model, a similar approach can also be followed. We may expect that  $\lambda_{L}\propto (\lambda_{'t\, Hooft}T)^{\alpha}$, where $\alpha$ is a constant that needs to be determined.  As we noted in Sec. \ref{DimenABJM}, we may take $\lambda_{'t\, Hooft}=\frac{N}{V_2}$, where $V_2$ is the volume of the two-dimensional space we have integrated over in going from $2+1$ to $0+1$ dimensions. From this definition of $\lambda_{'t\, Hooft}$, we see that it has the dimension of $[Length]^{-2}$, while we already know that each of $\lambda_L$ and $T$ have the dimension $[Length]^{-1}$. Putting these fact together, we have the equation for $\alpha$,
\be
[L]^{-1}=[L^{-2}L^{-1}]^{\alpha}=[L^{-3}]^{\alpha} \,,
\ee
and therefore we find $\alpha=1/3$. Thus, we may expect that $\lambda_{L}\propto (\lambda_{'t\, Hooft}T)^{1/3}$. In view of the equipartition theorem, this is consistent with the $\lambda_{L} \propto E^{1/3}$ based on the scaling symmetry as discussed in the previous section. Shortly, we will see the circumstances under which the remaining dimensionful parameter, namely, the mass, may effect the relation between the energy and temperature upon the application of the virial and the equipartition theorems. To prepare for the latter, let us first obtain the total number of independent degrees of freedom (d.o.f.) of the ABJM matrix model described by the action \eqref{SABJM}. Here, we have both $X_i$ and $\hat{X_i}$  as $N \times N$ Hermitian matrices with $i=1,2$. Each has $N^2$ real degrees of freedom, and therefore these give $4N^2$ d.o.f. in total. We also have the fields $Q^{\alpha}$ and $R^{\alpha}$ with $\alpha=1,2$ as $N \times N$ complex matrices, and each has $2N^2$ real d.o.f. leading to $8N^2$ real d.o.f. Therefore, the number of d.o.f. involved in the constituents of  \eqref{ABJMR2} is $4N^2+8N^2=12N^2$ before taking the global gauge symmetry and the constraints into account. Since the action is invariant under the $R$-symmetry group $SU(2)\times SU(2)\times U(1)\times U(1)\times Z_2$, each $SU(2)$ factor gives three and each $U(1)$ factor gives one, and thus in total eight real relations, while each of the equations in the Gauss-law constraint \eqref{gausslaw} gives $N^2$ real relations among  the unconstrained real degrees of freedom. Subtracting these from the latter, we find the independent d.o.f. count to be $12N^2-2N^2-8=10N^2-8$.

For the ansatz I,  in \eqref{ansatz11} there is a further reduction of the independent d.o.f., which comes about as follows. Since $X_i$ and $\hat{X_{i}}$ are null matrices due to vanishing of $\alpha(t)$ and $\beta(t)$ as the only admissible on-shell solution as argued in Sec. \ref{Ansatz1sec} and demonstrated in Appendix \ref{AppA}, we need to subtract out a factor of $4N^2$. Additionally,  because $R_{\alpha}=0$ in this ansatz, we need to subtract another factor of $4N^2$ d.o.f..  Finally, we also need to note that two equations of the Gauss-law constraint reduce to the same equation upon integrating by parts and taking the Hermitian conjugate of one or the other. Thus, the Gauss-law constraint imposes only $N^2$ real relations in this case. These facts bring the total number of d.o.f. count to $10N^2-8-8N^2+N^2=3N^2-8$. For large $N$, we may take $n_{d.o.f.}\approx 3N^2$. 

Let us now apply the virial theorem to the Hamiltonian in \eqref{hamiltonian1}.  Since the potential $V_N(\phi_1, \phi_2)$ is not a homogeneous polynomial of its arguments, there is no exact proportionality relation linking the average kinetic and potential energies. Instead, we have
\begin{align}
	2 \left \langle K\right \rangle& = 2 \left \langle V_N \right \rangle+2N^2(N-1)\left(\frac{8\pi\mu}{k} \phi_1^2\phi_2^2 +\frac{16\pi^2}{k^2}\phi_1^2\phi_2^4 +\frac{16\pi^2}{k^2}\phi_1^4\phi_2^2 \right)\nonumber \\
	&=: 2\left \langle V_N \right \rangle +\tilde{V}_N(\phi_1,\phi_2) \,,
\label{2K}
\end{align}
where the relevant expression in the first line provides the definition of $\tilde{V}_N(\phi_1,\phi_2)$ introduced in the second line. The latter  is positive definite for $k > 0$ (assuming that $\mu > 0$, too, indeed we set $\mu =1$), but this is not so for negative $k$. In fact, the minimum of $\tilde{V}_N(\phi_1,\phi_2)$ is given as 
\begin{align}\label{minv1}
	Min\left ( \tilde{V}_N(\phi_1,\phi_2)\right)=
	\begin{cases}
		0& \text{if both $k$ and $\mu$ have the same sign}\,, \\
		N^2(N-1) \frac{ 4 k\mu^3}{27\pi} & \text{if $k$ and $\mu$ have the opposite sign} \,.
	\end{cases}
\end{align}

Applying the equipartition theorem to the kinetic energy yields
\begin{align}\label{exkin1}
	\left \langle K\right \rangle = \frac{1}{2}(3N^2-8)T \approx \frac{3}{2} N^2 T \,,
\end{align}
where the approximation is valid at large $N$.

\vskip 1em

\textit{Case i: $k \geq1$: } 

\vskip 1em

In this case, $\tilde{V}_N(\phi_1,\phi_2)$  is positive definite, and therefore we have from \eqref{2K} the inequality $\left \langle K \right \rangle \geq \left \langle V_N \right \rangle$. This and \eqref{exkin1} together imply that $\langle E \rangle = \left \langle K \right \rangle + \left \langle V_N \right \rangle \leq n_{d.o.f} T \approx 3N^2 T$. We can express this inequality in the form 
\begin{equation}
	\frac{E}{N^2} \leq 3 T \,,
	\label{EvT}
\end{equation}
where we have also dropped the brackets on energy for ease in notation.

Since we expect that $\lambda_L \propto E^{1/3}$ due to the scaling properties of the model and also that  $\lambda_L \propto T^{1/3}$ as implied by the pure dimensional analysis, albeit both holding exact only in the massless limit, we are, nevertheless, led to examine the best-fitting curves of the form 
\begin{align}\label{fitform}
\lambda_L=\alpha_N\left(\frac{E}{N^2}\right)^{1/3} \,,
\end{align}
to profile the variation of the largest Lyapunov exponent as a function of $E/N^2$. The fitting curves are plotted in Fig. \eqref{fig1} for $k=1$, $N=5,10,15,20,25$, in Fig. \eqref{figk2}  for $k=2$ at the matrix levels $N=5,15,25$ and in Fig. \eqref{fign10} for $k=5,10$ at the matrix level $N=10$. We observe that these fits represent the variation of the largest Lyapunov exponent with respect to $E/N^2$ quite well. 

In fact, to evaluate the goodness of the fits in explaining the variation of $\lambda_L$ with respect to $E/N^2$, we may inspect the square of the multiple correlation coefficient, $R$ squared (we use $R_{sq}$ for short in what follows), and the residual sum of squares (SSE), which are usual statistical measures used for this purpose. The former takes a value between $0$ and $1$, with $R$ squared close to $1$ indicating better fits; i.e., a greater portion of the variance in the data is accounted for by the fitting curve, while the latter could take any positive value, with values close to zero indicating better fits. In the present context, $R_{sq}$ measures the correlation between the $\lambda_L$ values of the data and those predicted from the fitting curve, while SSE represents the total deviation of the predicted values from the fitting curve to the data. We find that for $k \geq 1$ the fits given in Figs. \eqref{fig1}, \eqref{figk2}, and  \eqref{fign10} have $R_{sq} \geq 0.97$ and with an average $R_{sq}  \approx 0.979$, i.e.,  the fitting curves accounting for the variation of the $\lambda_L$ with respect to $E/N^2$ around $ \approx 98 \%$ and average SSE values $\approx 0.43$. Coefficients of $\alpha_N$ for the fitting curves are provided Tables \ref{table:tablek1}, \ref{table:tablek2}, and \ref{table:tablen10}.  

\begin{table}[!htb]
	\centering
	\caption{$\alpha_N$ and $T_c$ values at $k=1$.}
	\begin{tabular}{| c | c | c | c | c | c | }
		\cline{2-6}
		\multicolumn{1}{c |}{} & $N=5$ & $N=10$ & $N=15$ & $N=20$ & $N=25$  \\  \hline 
		$\alpha_N$ &$0.9522$  & $0.713$  & $0.6092$ &$0.5448$  &$0.499$  \\ \hline 
		$T_c$ &$0.1022$  & $0.0662$  & $0.0523$ &$0.0442$  &$0.0388$  \\ \hline
	\end{tabular}
	\label{table:tablek1}
\end{table}

\begin{table}[!htb]
	\centering
	\caption{$\alpha_N$ and $T_c$ values at $k=2$.}
	\begin{tabular}{ | c | c | c | c | }
		\cline{2-4}
		\multicolumn{1}{c |}{}& $N=5$ & $N=15$ & $N=25$  \\  \hline 
		$\alpha_N$ &$0.7845$  & $0.4788$  & $0.3958$ \\ \hline 
		$T_c$ &$0.0764$  & $0.0364$  & $0.0274$\\ \hline
	\end{tabular}
	\label{table:tablek2}
\end{table}

\begin{table}[!htb]
	\centering
	\caption{$\alpha_N$ and $T_c$ values for $N=10$ at $ 5,10$.}
	\begin{tabular}{ | c | c | c | }
		\cline{2-3}
		\multicolumn{1}{c |}{} & $k=5$ & $k=10$  \\  \hline 
		$\alpha_N$  & $0.4168$  & $0.3338$ \\ \hline 
		$T_c$& $0.0296$  & $0.0212$\\ \hline
	\end{tabular}
	\label{table:tablen10}
\end{table}

We are now in a position to compare and relate our result to the MSS bound $\lambda_L \leq 2\pi T$ on the Largest Lyapunov exponent for quantum chaos \cite{Maldacena:2015waa}. This bound is conjectured to be satisfied in systems which are holographically dual to gravity, and it is shown in Ref. \cite{Maldacena:2016hyu} that it is saturated for the Sachdev-Ye-Kitaev fermionic matrix model. In Ref.  \cite{Gur-Ari:2015rcq}, chaotic dynamics of the BFSS matrix models are studied at the classical level, which provides an approximation to the quantum theory only in the high-temperature limit, and it was shown that the largest Lyapunov exponent disobeys the MSS bound only at sufficiently low temperatures. The authors of Ref. \cite{Gur-Ari:2015rcq} estimated the latter to be $\approx 0.015$. In Ref.  \cite{Baskan:2019qsb}, we study a deformation of the bosonic sector of the BFSS via two mass terms and investigating the chaos in this model via reduced effective Lagrangians, we were able to put upper bounds on the critical temperature above which MSS inequality is satisfied, and below which it will eventually be not obeyed. In Refs.  \cite{Buividovich:2017kfk, Buividovich:2018scl}, a so-called Gaussian state approximation(GSA) is introduced to investigate the quantum chaotic dynamics of the BFSS and related Yang-Mills matrix models. Results obtained in Refs.  \cite{Buividovich:2017kfk, Buividovich:2018scl} demonstrate that the largest and all the other Lyapunov exponents tend to zero at a nonzero value of the temperature and therefore comply completely with the MSS conjecture at all temperatures. Nevertheless, it remains an open problem to show if and how the BFSS model saturates the MSS bound.

As we noted in the Introduction, the ABJM model has a gravity dual \cite{Nastase:2015wjb} via the AdS/CFT correspondence. Therefore, we may expect the MSS conjecture to hold for quantum chaotic dynamics of the ABJM model too. In this article, we are investigating the dynamics of the mass-deformed ABJM model only at the classical level, as an approximation of the quantum theory in the high-temperature limit, so we should expect that the MSS bound eventually be not obeyed at sufficiently low temperatures.\footnote{Let us note in passing that the presence of mass terms may keep the system away from saturating the MSS bound even if the full quantum dynamics could be studied. However, mass deformations lead to non-trivial vacuum solutions in the form of fuzzy sphere matrix configurations and provide us a good departure point to probe the chaotic dynamics as we do in the present paper.} In other words, we expect the classical chaotic dynamics to comply with the MSS bound to a very large extent, while we also expect it to be insufficient to capture all the quantum features at low temperatures. Indeed, using  \eqref{EvT} and \eqref{fitform}, we find that there is a critical temperature, which we may denote as $T_c$ and is given by solving the equation
\be
\alpha_N(3T)^{1/3}=2\pi T_c \,,
\ee
which yields
\be
T_c = \sqrt{3}\left(\frac{\alpha_N}{2\pi}\right)^{3/2}.
\label{Tc}
\ee
From this result, we understand that for $T \geq T_c$ the present model complies with the MSS bound on $\lambda_L$, while for $T \leq T_c$, there is a temperature at and below which MSS bound is not respected. Thus, we may say that $T_c$ is an upper bound for the critical temperature at or below which MSS bound will eventually not be  obeyed by our model. The estimated $T_c$ values at the matrix levels $N=5,10,15,20,25$ are given in Tables \ref{table:tablek1} and \ref{table:tablek2} for $k=1$ and $k=2$, respectively, and in Table \ref{table:tablen10} at $N=10$ level  for $k=5,10$. We observe from the values of $T_c$ in these tables that with increasing matrix size, their values tend to decrease, which is in agreement with the fact that the 't Hooft limit is better emulated with increasing matrix size. From Table \ref{table:tablen10}, we also infer that $T_c$ values tend to decrease with increasing values of $k$; i.e., the models with larger CS coupling tend to comply with the MSS conjecture within a wider range of the temperature.

\vskip 1em

\textit{Case ii: $ k\leq-1$ :}

\vskip 1em

In this case, $\tilde{V}(\phi_1,\phi_2)_N$ is not positive definite as we have already noted; its minimum is negative and given by the expression in the second line of \eqref{minv1}. Adding and subtracting $\abs{Min(\tilde{V}_2)}$ to the \eqref{2K}, we may write
\begin{align}\label{virialtrick}
2 \langle K \rangle= 2 \langle V_N\rangle+\underbrace{\tilde{V}_N(\phi_1,\phi_2)+\abs{Min(\tilde{V}_N)}}_{\geq 0}-\abs{Min(\tilde{V}_N)},
\end{align}
which implies that
\begin{align}\label{KV2}
\langle K \rangle\geq\langle V_N \rangle-\frac{1}{2}\abs{Min(\tilde{V}_N)}\,.
\end{align}
We may therefore write
\begin{align}
E & =\underbrace{ \langle K \rangle+\langle V_N \rangle -\frac{1}{2}\abs{Min(\tilde{V}_N)}}_{\leq n_{d.o.f.}T}+\frac{1}{2}\abs{Min(\tilde{V}_N)} \,.
\end{align}
Using $\langle K \rangle \approx \frac{3}{2} N^2 T$ at large $N$, this leads to the inequality
\begin{align}
\frac{E}{N^2} - \gamma_N \leq  3T \,, \quad \gamma_N : = \frac{\abs{Min(\tilde{V}_N)}}{2N^2} \,. 
\label{Ineqansatz1}
\end{align}
In view of this relation, we conjecture to use best-fitting curves of the form 
\begin{align}\label{fitform2}
\lambda_ N=\alpha_N\left(\frac{E}{N^2}-\gamma_N\right)^{1/3} \,.
\end{align}
These curves are given in Fig. \eqref{fig-1} for $N=5,10,15,20,25$ at $k =-1$ and in Fig. \eqref{figk-2} for $N=10,15,25$ at $k=-2$. Similar to the previous case, they represent the variation of the largest Lyapunov exponent with respect to $E/N^2$ quite well, with $R_{sq} \geq 0.96$, with an average $R_{sq} \approx 0.972$ for the fitting curves in Figs. \eqref{fig-1} and \eqref{figk-2} and average SSE values $\approx 1.005$.

\begin{table}[!htb]
	\centering
	\caption{$\alpha_N$, $\gamma_N$, $T_c$ values at $k=-1$.}
	\begin{tabular}{| c | c | c | c | c | c | }
		\cline{2-6}
		\multicolumn{1}{c |}{} & $N=5$ & $N=10$ & $N=15$ & $N=20$ & $N=25$ \\  \hline 
		$\alpha_{N}$ &$0.8884$  & $0.633$  & $0.5529$ &$0.5018$  &$0.4648$  \\ \hline 
		$\gamma_{N}$ &$0.0943$  & $0.2122$  & $0.3301$ &$0.4480$  &$0.5659$ \\ \hline
		$T_c$ &$0.0920$  & $0.0533$  & $0.0452$ &$0.0390$  &$0.0348$ \\ \hline
	\end{tabular}
	\label{table:tablek-1}
\end{table}
\begin{table}[!htb]
	\centering
	\caption{$\alpha_N$, $\gamma_N$, $T_c$ values at $k=-2$.}	
	\begin{tabular}{| c | c | c | c | }
		\cline{2-4}
		\multicolumn{1}{c |}{} & $N=10$ & $N=15$ & $N=25$ \\  \hline 
		$\alpha_N$ &$0.4944$  & $0.4281$  & $0.357$   \\ \hline 
		$\gamma_N$ &$0.4244$  & $0.6602$  & $1.132$ \\ \hline
		$T_c$ &$0.0382$  & $0.0308$  & $0.0234$ \\ \hline
	\end{tabular}
	\label{table:tablek-2}
\end{table}

By the same line of reasoning discussed in the previous case, using \eqref{Ineqansatz1} and \eqref{fitform2}, we find that the critical temperature is given again as \eqref{Tc}, and the numerical estimates using the $\alpha_N$ values of the fitting curves at several different matrix levels are listed in Tables \ref{table:tablek-1} and \ref{table:tablek-2} for $k=-1$ and $k=-2$, respectively. 

Viewing the results of the cases \textit{i} and \textit{ii} together, we conclude that $T_c$ values decrease with increasing $N$ and/or $\abs{k}$; i.e., the MSS bound is respected in a wider range of the temperature at matrix levels which better capture the 't Hooft limit and/or at larger values of the CS coupling. 

\section{Ansatz II}
\label{AnsatzIIsec}

We would like to introduce another ansatz configuration with nonzero $R_{\alpha}$ and $Q_{\alpha}$ matrices and examine the ensuing dynamics. We consider the ansatz
\begin{align}\label{AnsatzII}
	Q_1 &= q(t)G_1 \,, \quad R_1 = r(t)G_1\,,  \nn \\
	Q_2 &= q(t)G_2 \,, \quad R_2 = r(t)G_2 \,,
\end{align}
while we still take $X_i$ and $\hat{X_i}$ as arbitrary diagonal matrices as given in \eqref{ansatz11}. This configuration satisfies the Gauss-law constraints given in \eqref{gausslaw}, as can easily be checked, and the equations of motion for $\alpha(t)$ and $\beta(t)$ yield the only real solution as the trivial solution $\alpha(t)=\beta(t)=0$ as shown in Appendix \ref{AppA}. Thus, in this case, too, we set $X_i = 0 = \hat{X}_i$ in what follows.

Substituting the matrix configuration \eqref{AnsatzII} into the action \eqref{ABJMR2}, and performing the trace over the GRVV matrices, we obtain the effective Lagrangian as
\begin{multline}
L_N(q(t) \,, r(t)) = N^2(N-1) \Big ( \dot{q}^2 + \dot{r}^2 - \mu^2q^2 - \mu^2r^2- \frac{8\pi\mu}{k}q^4 + \frac{8\pi \mu}{k}r^4 \\
+ \frac{12\pi^2}{k^2}q^4 r^2+ \frac{12\pi^2}{k^2}q^2 r^4 -\frac{16\pi^2}{k^2}q^6 - \frac{16\pi^2}{k^2}r^6 \Big) \,.
\end{multline}
The corresponding Hamiltonian is
\begin{align}
	\begin{split}\label{ham2}
		H_N(q(t) \,, r(t)) &= \frac{p_q^2}{4N^2(N-1)} + \frac{p_r^2}{4N^2(N-1)} + N^2(N-1)\Big(\mu^2q^2 +\mu^2r^2+ \frac{8\pi \mu}{k}q^4\\
		& \quad \quad \quad - \frac{8\pi\mu}{k}r^4 - \frac{12\pi^2}{k^2}q^4 r^2- \frac{12\pi^2}{k^2}q^2 r^4 + \frac{16\pi^2}{k^2}q^6 + \frac{16\pi^2}{k^2}r^6\Big) \\
		&:=\frac{p_{q}^2}{4N^2(N-1)}+\frac{p_{r}^2}{4N^2(N-1)}+V_N(q,r)\,,
	\end{split}
\end{align}
where $V_N(q,r)$ is the effective potential defined by the relevant terms in the first two lines of \eqref{ham2}. A few remarks regarding the structure of  $V_N(q,r)$ are now in order. Let us first note that this potential is not positive definite for either $k > 0$ or $k < 0$, while its minimum is at zero. Next, we easily see that $V_N(q,r)$ is symmetric under the exchange of $q$ and $r$. For $k\leftrightarrow -k$, the two terms which are proportional to $\frac{1}{k}$ change sign, but this can be compensated by exchanging $q$ and $r$. Thus, we conclude that the dynamics due to this potential is independent of the sign of $k$. 

Hamilton's equations of motion are easily obtained and are given below: 
\begin{subequations}
	\begin{align}
		&\dot{q}-\frac{p_{q}}{N^2(N-1)} =0 \,, \\
		&\dot{r}-\frac{p_{r}}{N^2(N-1)}=0 \,, \\
		&\dot{p_{q}}+N^2(N-1)\left(2\mu^2q+\frac{32\pi \mu}{k}q^3-\frac{48\pi^2}{k^2}q^3 r^2 -\frac{24\pi^2}{k^2}qr^4 + \frac{96\pi^2}{k^2}q^5\right)=0 \,, \\
		&\dot{p_{r}}+N^2(N-1)\left(2\mu^2r- \frac{32\pi \mu}{k}r^3 - \frac{24\pi^2}{k^2}q^4 r - \frac{48\pi^2}{k^2}q^2 r^3 + \frac{96\pi^2}{k^2}r^5\right)=0 \,. 
	\end{align}
\label{Heom2}
\end{subequations}

To gain more insight about this Hamiltonian system, we explore its fixed points and their stability at the linear order. The details of  this analysis are relegated to Appendix \ref{AppB}. We find that for real values of $\mu$ either the set $(0,\pm\frac{\sqrt{k \mu}}{2\sqrt{3\pi}},0,0)$ or the set $(\pm\frac{\sqrt{- k \mu}}{2\sqrt{3\pi}},0,0,0)$ gives unstable fixed points for $k \mu > 0$ and $k \mu < 0$,respectively, while the remaining fixed points are of borderline type. The corresponding energies in either case are
\be
E_F = N^2 (N-1) \frac{ | k \mu ^3 |}{27 \pi } \,,
\label{EF}
\ee
and the system is likely to exhibit dynamical evolution which is chaotic at and above these energies. This suggests that we may consider an offset $\gamma_N^{(1)} := E_F/N^2 = (N-1) \frac{ | k \mu ^3|}{27 \pi }$ for the fitting curves of the form $\lambda_L = \alpha_N (\frac{E}{N^2} - \gamma_N)^{1/3}$. In the next subsection, we compare this with the values of $\gamma_N$ implied upon the use of the virial and equipartion theorems.

\subsection{Dependence $\lambda_L$ on energy and temperature}\label{Tdependence2}

To obtain the profile of the mean largest Lyapunov exponent $\lambda_L$ with respect to the variation of $E/N^2$, we numerically solve the Hamilton equations \eqref{Heom2} and evaluate the mean of $\lambda_L$ by averaging out the largest Lyapunov exponents over $100$ runs of the code with randomly selected initial conditions. For this ansatz, numerical aspects of the initial condition selection turn out to be somewhat more conveniently handled by setting  $q(0)=0$. Using three random numbers $\omega_i$ $(i=1,2,3)$  and writing  
$\Omega_i = \frac{\omega_i}{\sqrt{\omega_i^2}}\sqrt{E}$ as in the case of ansatz I, we generate the initial conditions in the form
\begin{align}
p_q(0) =  \pm 2 N \sqrt{(N-1)} \Omega_1 \,, \quad p_r(0) = \pm 2 N \sqrt{(N-1)} \Omega_2 \,, \quad V_N(q(0)=0 \,, r(0)) = \Omega_3^2 \,,
\label{ic}
\end{align}
where the last equation in \eqref{ic} takes the explicit form
\begin{align}\label{initial3}
	N^2(N-1)\left(\mu^2r(0)^2- \frac{8\pi \mu}{k}r(0)^4+\frac{16\pi^2}{k^2}r(0)^6\right)-\Omega_3^2 = 0 \,,
\end{align}
and its real roots are used to pick $r(t)$ at $t=0$, i.e., the $r(0)$ value.

Applying the virial theorem, we find that 
\be\label{virial2}
2 \langle K \rangle = 2 \langle V_N \rangle+ \tilde{V}_N(q,r) \,,
\ee
where
\be
 \tilde{V}_N(q,r) = N^2(N-1) \bigg(\frac{16\pi \mu}{k}q^4- \frac{16\pi \mu}{k}r^4-\frac{48\pi^2}{k^2}q^4 r^2 -\frac{48\pi^2}{k^2}q^2 r^4
 + \frac{64\pi^2}{k^2}q^6 + \frac{64\pi^2}{k^2}r^6 \bigg) \,.
\ee
Evaluating the minimum of $\tilde{V}_N(q,r)$, we find that it is given as 
\begin{align}
Min(\tilde{V}_N(q,r)) = - \frac{N^2(N-1)64\abs{k\mu^3}}{135\sqrt{5}\pi} \,.
\end{align}
Following the same line of development and steps as in Sec. \ref{Tdependence}, we have
\begin{align}\label{EN2dof}
\frac{E}{N^2} -\gamma_N\leq \frac{n_{d.o.f.}T}{N^2} \,, \quad \gamma_N^{(2)} :=  \frac{\abs{Min(\tilde{V}_N(q,r))}}{2N^2}  \,.
\end{align}
From this consideration as well as the energies of the unstable fixed points, we are led to consider best-fitting curves of the form 
\begin{align}\label{fit2}
	\lambda_L=\beta_N \left ( \frac{E}{N^2} - \gamma_N \right)^{1/3}
\end{align}
to the $\lambda_L$ versus $E/N^2$ data. Values of  $\gamma_N^{(1)}$ and  $\gamma_N^{(2)}$ are comparable, and in what follows we use the latter as they tend to work slightly better with the fitting curves.

Let us recall once again that, depending on the structure of the potential, chaos and quasiperiodic motion can coexist and may fill comparable hypervolumes of the phase space at a given energy. For the model emerging from ansatz II, we also let our code check if the largest Lyapunov exponent has a value below a certain threshold at the final time (here, we continue to use a time step of $0.25$ and run the code from time $0$ to $3000$) and do not include it in the averaging over the initial conditions. From numerics, we found that roughly $\approx 1/5$ to $\approx 1/10$ of the initial conditions lead to quasiperiodic orbits at low energies, but their number too also tends to zero with increasing energy. Applying this process allows us to evaluate the average $\lambda_L$ value at a given energy with high precision, which is otherwise only obtained with relatively large root-mean-square errors. In our computations, we picked this threshold as $ 0.05$ after a number of numerical trials.\footnote{Except for the cases $| k | = 2 \,, (N=25$), $| k | = 5 \,,10 \,, (N =10) $ for which we picked the thresholds as $0.075\,, 0.1$ and $0.2$, respectively.}

\begin{figure}[!htb]
	\begin{subfigure}[!htb]{.5\textwidth}
		\centering
		\includegraphics[width=6.6cm]{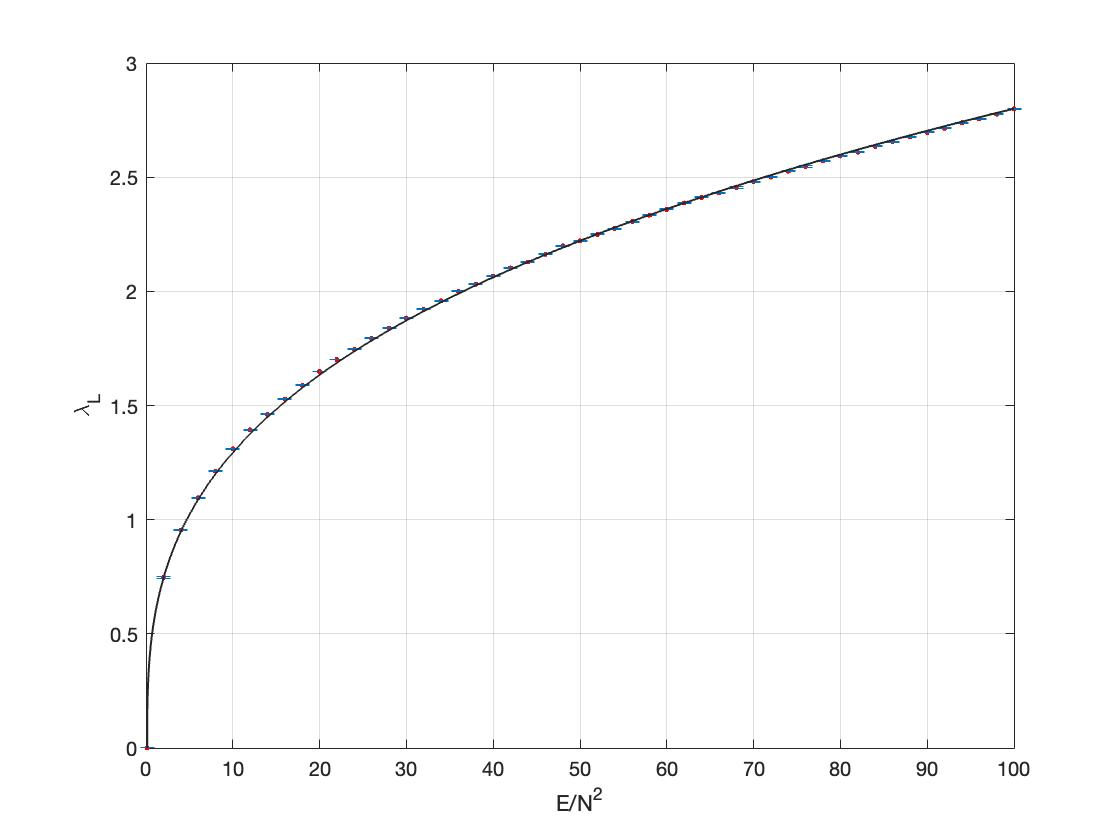}
		\caption{$N=5$}
		\label{fig:lenn5k-1}
	\end{subfigure}
	\begin{subfigure}[!htb]{.5\textwidth}
		\centering
		\includegraphics[width=6.6cm]{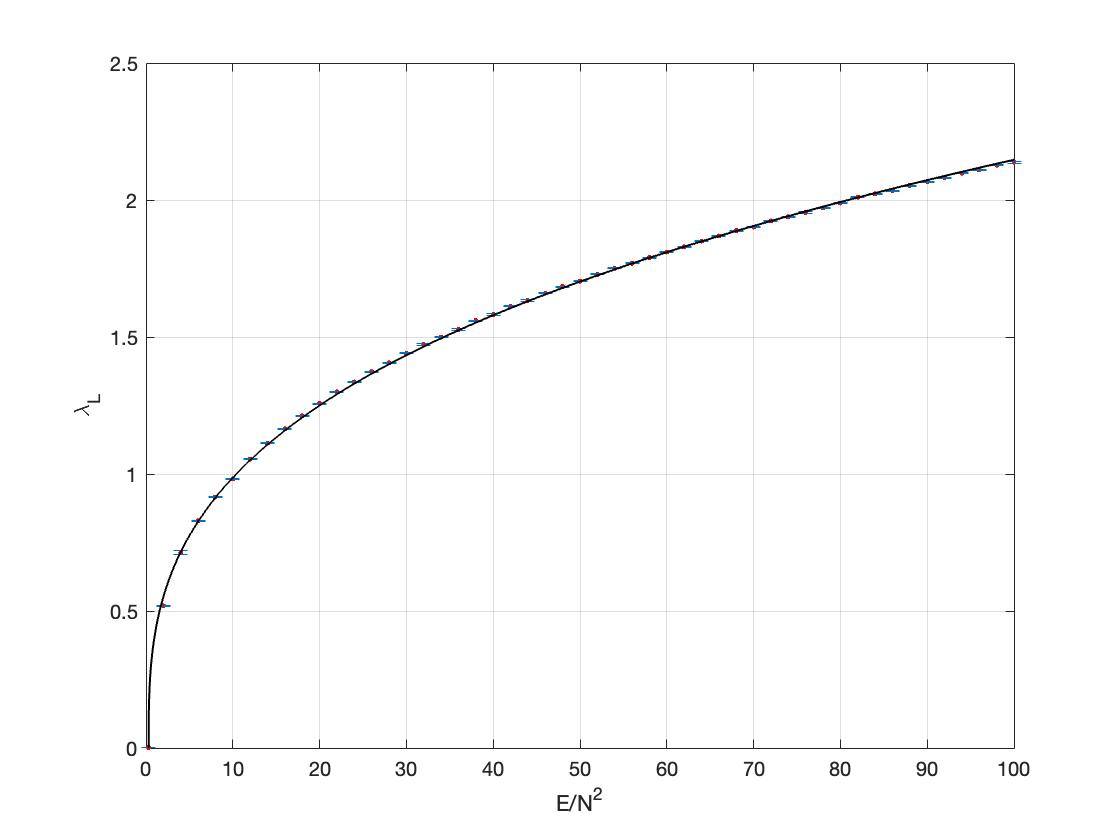}
		\caption{$N=10$}
		\label{fig:lenn10k-1}
	\end{subfigure}
	\begin{subfigure}[!htb]{.5\textwidth}
		\centering
		\includegraphics[width=6.6cm]{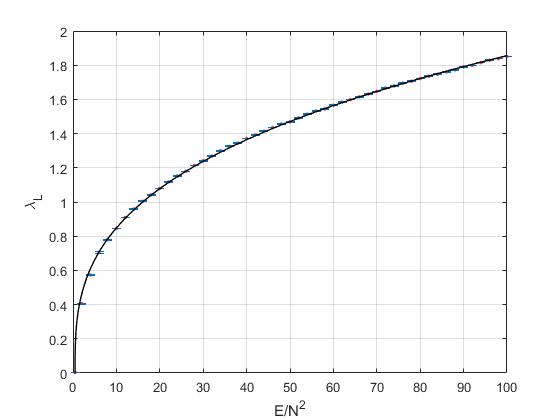}
		\caption{$N=15$}
		\label{fig:lenn15k-1}
	\end{subfigure}
	\begin{subfigure}[!htb]{.5\textwidth}
		\centering
		\includegraphics[width=6.6cm]{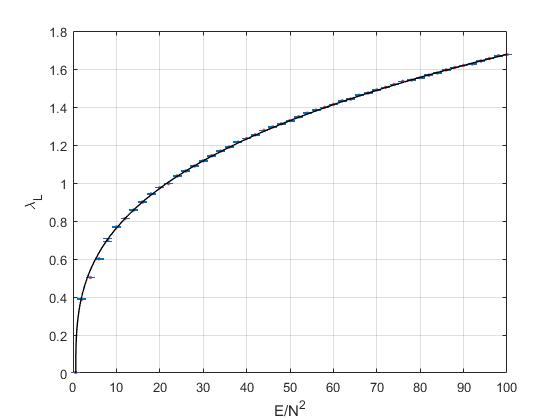}
		\caption{$N=20$}
		\label{fig:lenn20k-1}
	\end{subfigure}
	\begin{center}
		\begin{subfigure}[!htb]{.5\textwidth}
			\centering
			\includegraphics[width=6.6cm]{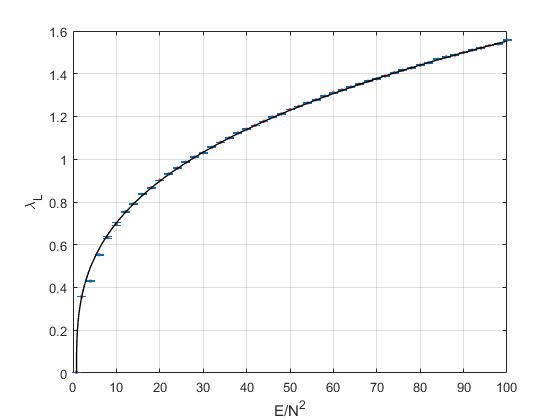}
			\caption{$N=25$}
			\label{fig:lenn25k-1}
		\end{subfigure}
	\end{center}
	\caption{Largest Lyapunov exponent and the best-fitting curves in the form $\lambda_L=\beta_N(\frac{E}{N^2}-\gamma_N)^{1/3}$ at $k= \pm1$.}
	\label{fig2k-1}
\end{figure}

For all cases of interest, it appears sufficient to use $E/N^2$ in the range $(0,100)$. For $ | k | = 1$, the data points and the fitting curves are depicted in Fig. \ref{fig2k-1} for the matrix levels $N=5,10,15,20,25$ and the $\beta_N$ coefficients of the fits are given in Table \ref{table:table2k-1}. 

\begin{table}[!htb]
	\centering
	\caption{$\beta_N$, $\gamma_N$ and $T_c$ values at $k=\pm1$.}	
	\begin{tabular}{ | c | c | c | c | c | c | }
		\cline{2-6}
		\multicolumn{1}{c |}{} & $N=5$ & $N=10$ & $N=15$ & $N=20$ & $N=25$ \\  \hline 
		$\beta_N$ &$0.6035$  & $0.463$  & $0.4005$ &$0.3621$  &$0.3355$\\ \hline 
		$\gamma_N$ &$0.1350$  & $0.3037$  & $0.4724$ &$0.6411$  &$0.8098$  \\ \hline
		$T_c$ &$0.0787$  & $0.0529$  & $0.0425$ &$0.0366$  &$0.0326$  \\ \hline
	\end{tabular}
	\label{table:table2k-1}
\end{table}

\begin{figure}[!htb]
	\begin{subfigure}[!htb]{.5\textwidth}
		\centering
		\includegraphics[width=6.6cm]{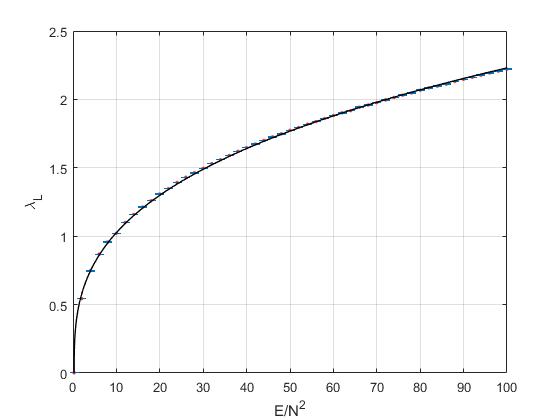}
		\caption{$N=5$}
		\label{fig:lenn5k-2}
	\end{subfigure}
	\begin{subfigure}[!htb]{.5\textwidth}
		\centering
		\includegraphics[width=6.6cm]{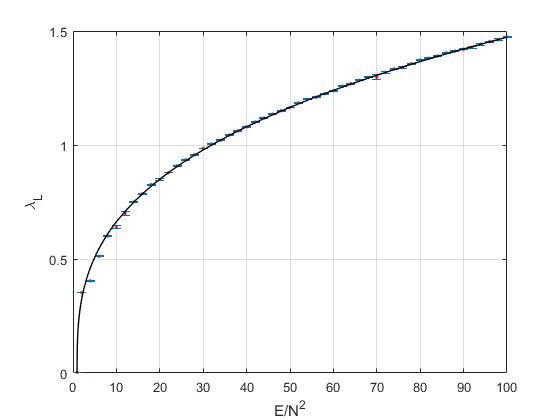}
		\caption{$N=15$}
		\label{fig:lenn15k-2}
	\end{subfigure}
	\begin{subfigure}[!htb]{.5\textwidth}
		\centering
		\includegraphics[width=6.6cm]{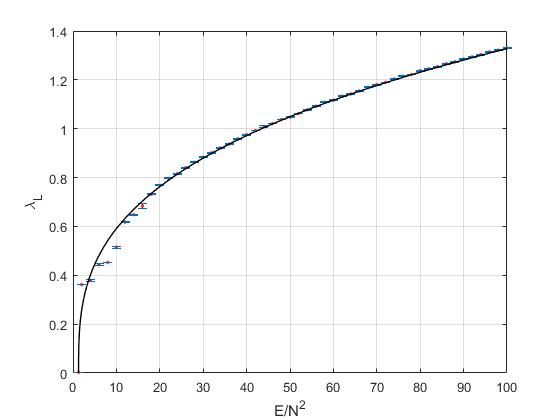}
		\caption{$N=20$}
		\label{fig:lenn20k-2}
	\end{subfigure}
	\begin{subfigure}[!htb]{.5\textwidth}
		\centering
		\includegraphics[width=6.6cm]{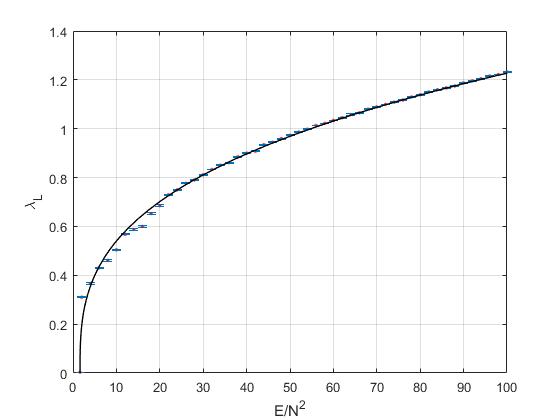}
		\caption{$N=25$}
		\label{fig:lenn25k-2}
	\end{subfigure}
	\caption{Largest Lyapunov exponent and the best-fitting curves in the form $\lambda_L=\beta_N(\frac{E}{N^2}-\gamma_N)^{1/3}$ at $k= \pm2$.}
	\label{fig2k-2}
\end{figure}

To profile the variation of $\lambda_L$ at larger values of CS coupling, we first inspect the case $k= \pm 2$. Data points and fitting curves are given in Fig. \eqref{fig2k-2}, and the corresponding $\beta_N$ values are listed in Table \ref{table:table2k-2}. 

\begin{table}[!htb]
	\centering
	\caption{$\beta_N$, $\gamma_N$ and $T_c$ values at $k=\pm2$.}
	\begin{tabular}{ | c | c | c | c | c | }
		\cline{2-5}
		\multicolumn{1}{c |}{} & $N=5$ & $N=15$ & $N=20$ & $N=25$ \\  \hline 
		$\beta_N$ &$0.4809$  & $0.3183$  & $0.2873$ & $0.2659$ \\ \hline 
		$\gamma_N$ &$0.2699$  & $0.9448$  & $1.2822$  & $1.620$  \\ \hline
		$T_c$ &$0.0560$  & $0.0302$  & $0.0258$ & $0.0230$  \\ \hline
	\end{tabular}
	\label{table:table2k-2}
\end{table}

\begin{figure}[!htb]
	\begin{subfigure}[!htb]{.5\textwidth}
		\centering
		\includegraphics[width=6.6cm]{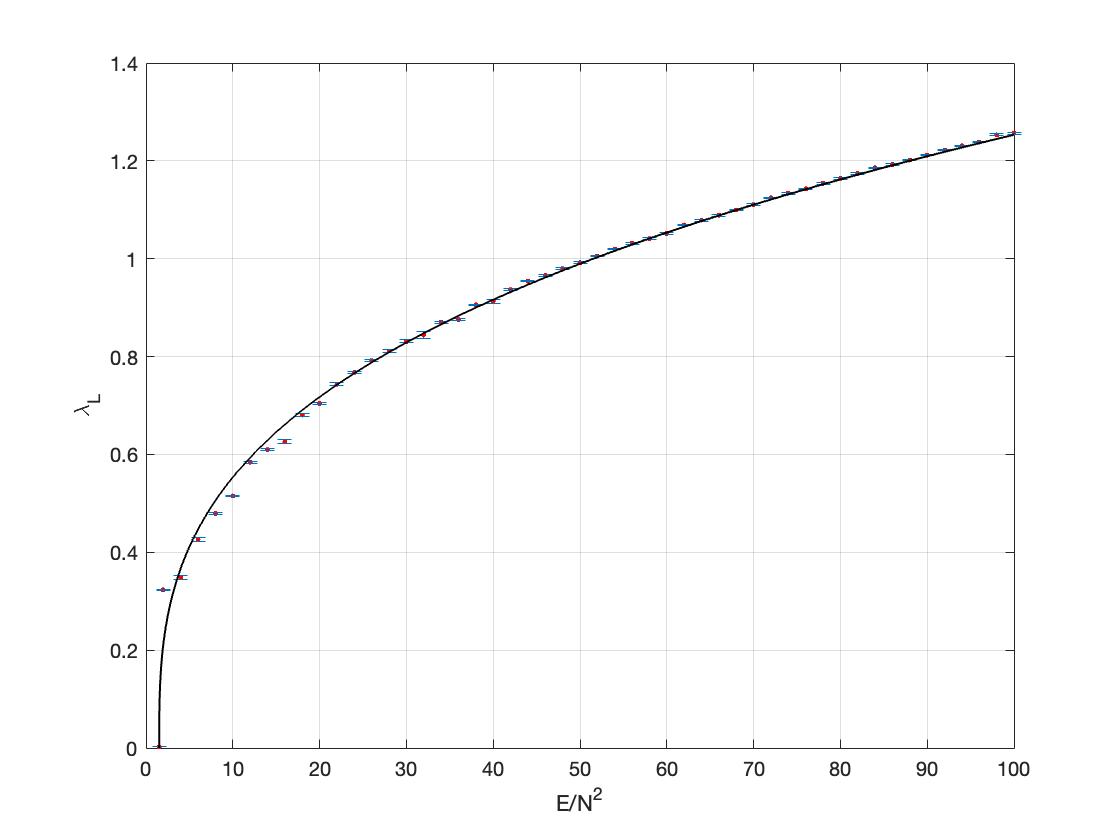}
		\caption{$k= \pm 5$}
		\label{fig:n10k52}
	\end{subfigure}
	\begin{subfigure}[!htb]{.5\textwidth}
		\centering
		\includegraphics[width=6.6cm]{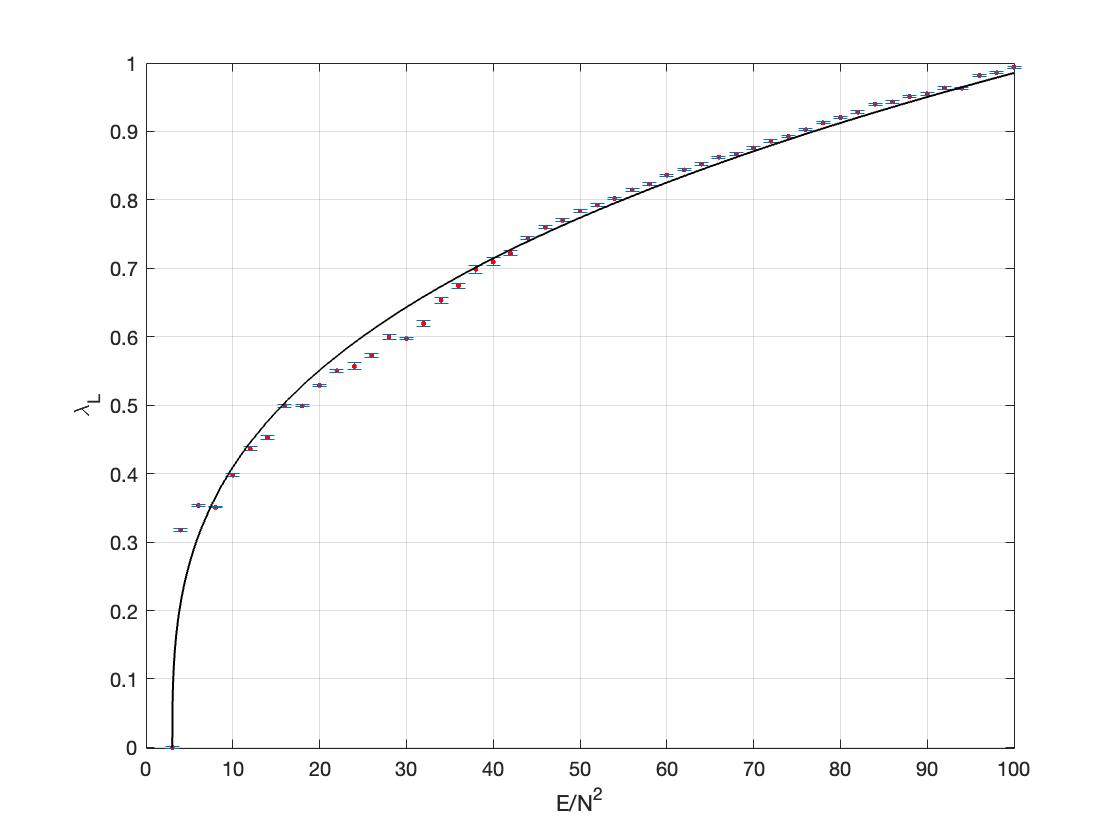}
		\caption{$k= \pm 10$}
		\label{fig:n10k102}
	\end{subfigure}
	\caption{Largest Lyapunov exponent and the best-fitting curves in the form $\lambda_L=\beta_N(\frac{E}{N^2}-\gamma_N)^{1/3}$ for $N=10$ at $k= \pm 5, \pm10$.}
	\label{n10k2}
\end{figure}

At $k=\pm5 \,, \pm10$ the data for $\lambda_L$ and the corresponding best-fitting curves are provided in Fig. \ref{n10k2} with $\beta_N$ coefficients listed in Table \ref{table:n10k102}. 

\begin{table}[!htb]
	\centering
	\caption{$\beta_N$, $\gamma_N$ values at $N=10$ at $k= \pm 1, \pm 5, \pm 10$.}	
	\begin{tabular}{ | c | c | c | c | }
		\cline{2-4}
		\multicolumn{1}{c |}{} & $k= \pm 1$ & $k= \pm 5$ & $k= \pm 10$ \\  \hline 
		$\beta_N$ &$0.463$  & $0.2714$  & $0.2145$ \\ \hline 
		$\gamma_N$ &$0.3037$  & $1.5184$  & $3.0369$\\ \hline
		$T_c$ &$0.0529$  & $0.0237$  & $0.0167$  \\ \hline
	\end{tabular}
	\label{table:n10k102}
\end{table}

From the plots provided in Figs. \ref{fig2k-1}, \ref{fig2k-2}, and \ref{n10k2}, we observe that the fitting curves represent the Lyapunov data almost perfectly. We find that all the fitting curves given in these figures $R_{sq} \geq 0.99$, while the SSE values generally vary around $\approx 0.002$ to $\approx 0.007$, except for a few cases ($k= \pm 2$, $N=25$, and $k = \pm 5 \,, \pm 10$, $N=10$) for which they are around $\approx 0.02$. Thus the fitting curves capture the variation of the $\lambda_L$ with respect to $E/N^2$ around $\approx 99 \%$. 

To obtain critical upper bound temperatures, $T_c$, using the coefficients $\beta_N$ of the fitting curves, we need to count the independent degrees of freedom of the matrices given in \eqref{AnsatzII}. In contrast to our first ansatz \eqref{ansatz11}, in this case, $R_{\alpha}$ matrices are no longer zero. Thus, we need to note that $R_{\alpha}$ ($\alpha =1,2$) contribute $4N^2$ degrees of freedom in total, while the two equations of the Gauss-law constraint \eqref{gausslaw} imply the same condition upon integration by parts of one or the other equation. Thus, the Gauss-law imposes only $N^2$ real constraints in this case, too. We therefore have $n_{d.o.f.} = 7N^2 - 8$, which in the large-$N$ limit is given as $n_{d.o.f.}\approx7N^2$. Therefore, Eq. \eqref{EN2dof} immediately leads to the inequality
\begin{align}
\frac{E}{N^2}-\gamma_N\leq7T \,.
\end{align}
We find that the critical temperature is obtained by solving
\begin{align}
\beta_N(7T)^{1/3}=2\pi T  \,,
\end{align}
and this yields 
\be
T_c = \sqrt{7}(\frac{\beta_N}{2\pi})^{3/2} \,.
\ee
Our estimates for the critical temperatures are given in Tables \ref{table:table2k-1}, \ref{table:table2k-2}, and \ref{table:n10k102}. Let us note that classical chaotic dynamics of the family of effective Hamiltonians, $H_N$, comply with the MSS bound for $T > T_c$, while they will eventually not obey it at or below $T_c$ values. Similar to the result obtained for ansatz I, we notice that with increasing matrix size and/or CS coupling values $k$ critical temperatures decrease. In particular, it is interesting to note that $T_c \approx 0.0167$ for $N=10$ and $k= \pm 10$, which is comparably close to $\approx 0.015$ found in Ref. \cite{Gur-Ari:2015rcq} for the BFSS model, although the two models are quite different in terms of the power-law dependence of $\lambda_L$s on energy ($\propto E^{1/3}$ for the ABJM model and $\propto E^{1/4}$ for the BFSS model). 

\section{Conclusions and Outlook}
\label{con}

In this paper, we performed a detailed study of the chaotic dynamics of the mass-deformed ABJM model. Working in the 't Hooft limit, and assuming that all the fields are spatially uniform and introducing ansatz configurations involving fuzzy spheres in the form of GRVV matrices with collective time dependence, we have obtained effective models and computed their Lyapunov exponents using numerical algorithms. Our results clearly indicate that these models possess chaotic dynamics. In particular, we directed our attention to the profile of the largest Lyapunov exponent and found that, depending on the form of the effective potential, 
either $\lambda_L \propto (E/N^2)^{1/3}$ or $\lambda_L \propto (E/N^2 - \gamma_N)^{1/3}$, where $\gamma_N(k, \mu)$ is a constant determined in terms of the Chern-Simons coupling $k$, the mass $\mu$, and the matrix level $N$. They represent the result of the numerical findings considerably well as it is observed from Figs. \eqref{fig1}--\eqref{n10k2} and also further corroborated by the $\lambda_L \propto E^{1/3}$ power-law dependence due to the scaling symmetry of the model in the massless limit. Upon the use of the virial and the equipartition theorems, we were able to examine the temperature dependence of the $\lambda_L$'s and derived critical upper bounds, $T_c$, on the temperature above which the MSS inequality, $\lambda_L \leq 2 \pi T $, is respected and below which it will eventually not be obeyed. Our numerical finding for these $T_c$ values are presented in the tables given in Secs. \ref{Ansatz1sec} and \ref{AnsatzIIsec}, from which it is also observed that the $T_c$ values display a decreasing trend with increasing matrix size, i.e., with the better numerical emulation of the 't Hooft limit, as well as with the increasing values of the CS coupling $k$.   

We strongly feel that the next step is to devise new methods to go beyond the classical analysis presented in the present paper and explore the quantum dynamics of these models. The latter appears to be quite a formidable task. Nevertheless, inspired by the methods used in quantum chemistry in approaching many-body problems, recently a new real-time method,\footnote{Most of the earlier investigations as well as some recent studies \cite{Kawahara:2006hs, Kawahara:2007fn, DelgadilloBlando:2007vx, DelgadilloBlando:2008vi, Asano:2018nol} have been aimed at investigating the phase structure of these models in the Euclidean time formulation using both analytical and Monte-Carlo methods.} which can be named the Gaussian state approximation, was developed and thoroughly applied to the BFSS model \cite{Buividovich:2018scl,Buividovich:2017kfk}. In its simplest form, GSA aims at incorporating the quantum corrections by considering a larger but a truncated set of observable whose Heisenberg equations of motion are obtained via the use of a Gaussian density matrix. Application of this method to the BFSS model demonstrated that all the Lyapunov exponents tend to zero at a nonvanishing temperature, implying that the quantum description of the BFSS model within the GSA approximation is fully compliant with the MSS inequality. However, given that it is still only an approximation of the full quantum dynamics, it falls short of providing an explicit saturation of the MSS bound by the largest Lyapunov exponent, in contrast to the result for the Sachdev-Ye-Kitaev model obtained in Ref. \cite{Maldacena:2016hyu} and expected for all models with holographic duals according to the MSS conjecture. We think that it will be extremely useful to attempt to apply the GSA to the ABJM model as well to the family of effective Hamiltonians introduced in the present manuscript, not only to test the usefulness of GSA beyond the BFSS model but also to probe the quantum chaotic dynamics of the ABJM model. We hope to report on the possible developments along this direction elsewhere. 

\section*{Acknowledgement}
Authors acknowledge the support of TÜBİTAK under the Project No. 118F100.

\appendix

\section{Equations of motion for $\alpha(t)$ and $\beta(t)$}
\label{AppA}

\subsection{Ansatz I}

With the ansatz configuration given as 
\begin{align}
	\begin{split}
		X_i &= \alpha(t)\text{diag}((A_i)_1,(A_i)_2,...,(A_i)_N)  : =  \alpha(t) \, A_i \,, \\
		\hat{X}_i &= \beta(t)\text{diag}((B_i)_1,(B_i)_2,...,(B_i)_N) : =  \beta(t) \,  B_i \,, \\
		Q_\alpha &= \phi_\alpha(t)G_\alpha, \quad R_\alpha=0 \,, 
	\end{split}
\end{align}
and working in the $A_0 =0$, ${\hat A}_0 = 0$ gauge, we immediately see that CS part of the action vanishes identically:
\beqa
& - & \frac{k}{4\pi} \Tr(\epsilon^{ij} X_i  \dot{X}_j) + \frac{k}{4\pi}\Tr(\epsilon^{ij} \hat{X}_i  \dot{\hat{X}}_j ) \,, \nn \\
& = & - \frac{k}{4\pi} \alpha \dot{\alpha} \Tr \lbrack A_1 \,, A_2 \rbrack + \frac{k}{4\pi} \beta \dot{\beta} \Tr \lbrack B_1 \,, B_2 \rbrack \,, \nn \\
&=& 0 \,.
\eeqa
Next, we evaluate $\Tr\abs{D_i Q_1}^2+\Tr\abs{D_i Q_2}^2$. We have 
\begin{align}
(D_i Q_{\alpha})_{ab}&=i(X_i Q_{\alpha})_{ab}-i(Q_{\alpha}\hat{X_i})_{ab} \,, \\ \nonumber
&=i(X_i)_{ac}(Q_{\alpha})_{cb}-i(Q_{\alpha})_{ac}(Q_{\alpha})_{cb} \,,
\end{align}
where the indices $a,b: 1\,, \cdots \,, N$. Using \eqref{GRVValb}, we obtain, for $\alpha =1$,
\begin{align}
(D_i Q_{1})_{ab}&=i(X_i Q_{1})_{ab}-i(Q_{1}\hat{X_i})_{ab} \,, \nonumber \\
&=i(X_i)_{ac}(\phi_1 G_1)_{cb}-i(\phi_1 G_1)_{ac}(\hat{X_i})_{cb} \,, \nonumber \\
&=i\alpha(t)\phi_1(t)(A_i)_{ac}\sqrt{c-1}\delta_{cb}-i\beta(t)\phi_1(t)\sqrt{a-1}\delta_{ac}(B_i)_{cb} \,, \nonumber \\
&=i\alpha \phi_1 \sqrt{b-1}(A_i)_{ab}-i\beta\phi_1\sqrt{a-1}(B_i)_{ab} \,, \nonumber \\
&=i\phi_1(\sqrt{b-1}X_i-\sqrt{a-1}\hat{X_i})_{ab} \,, \nonumber \\ 
&=i \phi_1 \sqrt{a-1} (X_i -\hat{X_i})_{ab} \,,
\end{align}
where the last line follows since $X_i$ and $\hat{X_i}$ are diagonal. The corresponding Hermitian conjugate is 
\begin{align}
{(D_i Q_{\alpha})}_{ab}^{\dagger} = - i \phi_1 \sqrt{a-1} (X_i -\hat{X_i})_{ba} \,.
\end{align}
These give 
\begin{align}
	\Tr\abs{D_iQ_1}^2 &=Tr(D_iQ_1)^{\dagger}(D_iQ_1)={(D_i Q_1)^{\dagger}}_{ab}(D_i Q)_{ba \,,}  \nonumber  \\
	&=\phi_1^2 (a-1) \left ((X_i^2)_{aa}+ (\hat{X_i}^2)_{aa} - 2 (X_i \hat{X_i})_{aa} \right ) \,, \nonumber \\
	&={\phi_1}^{2} \sum_{\substack{a=1 \\ i=1,2}}^{N} (a -1) \lbrack \alpha^2  (A_a^i)^2 + \beta^2  (B_a^i)^2 -2 \alpha \beta  A_a^i  B_a^i \rbrack \,, \nonumber \\
	&= {\phi_1}^{2} \sum_{\substack{a=1 \\ i=1,2}}^{N} (a -1) ( \alpha A_a^i - \beta B_a^i)^2 \,.
\end{align}
Similarly, for $\alpha =2$,
\begin{align}
(D_i Q_2)_{ab}&=i(X_i)_{ac}\phi_2(G_2)_{cb}-i(\phi_2 G_2)_{ac}(\hat{X_i})_{cb} \,, \\  \nonumber
&=i\phi_2\sqrt{N-c}\,(X_i)_{ac}\delta_{c+1,b}-i\phi_2\sqrt{N-a}\,\delta_{a+1,c}(\hat{X_i})_{cb} \,, \\  \nonumber
&=i\phi_2(\sqrt{N-b+1}\,(X_i)_{a,b-1}-\sqrt{N-a}\,(\hat{X_i})_{a+1,b} ) \,, \nonumber \\
&= i \phi_2 \sqrt{N-a} ((X_i)_{a,b-1} - (\hat{X_i})_{a+1,b}) \,,
\end{align}
with the Hermitian conjugate given as
\begin{align}
(D_i Q_2)_{ab}^{\dagger}=-i\phi_2 \sqrt{N-a} ((X_i)_{b-1,a} - (\hat{X_i})_{b,a+1}) \,.
\end{align}
These give 
\begin{align}
\Tr\abs{D_i Q_2}^2 &= \phi_2^2 (N-a) \left( (X_i^2)_{a-1,a-1}+ (\hat{X_i})_{a+1,a+1} - (X_i)_{a,b-1}(\hat{X_i})_{b,a+1}- (\hat{X_i})_{a+1,b}(X_i)_{a,b-1}\right)	\,, \nonumber \\
& = \phi_2^2 \sum_{\substack{a=1 \\ i=1,2}}^{N}  (N-a) \lbrack \alpha^2  (A_{a-1}^i)^2 + \beta^2  (B_{a+1}^i)^2 - 2 \alpha \beta  A_a^i  B_{a+1}^i  \rbrack \,.
\end{align}
We therefore have,
\begin{align}
\Tr\abs{D_i Q_1}^2+\Tr\abs{D_i Q_2}^2=\alpha^2 S_1 +\beta^2 S_2 - 2\alpha\beta S_3 \,, 
\end{align}
where
\begin{align}
S_1 & = {\phi}_1^{2} \sum_{\substack{a=1 \\ i=1,2}}^{N} (a -1) (A_a^i)^2 +  \phi_2^2 \sum_{\substack{a=1 \\ i=1,2}}^{N}  (N-a)  (A_{a-1}^i)^2 \,, \nonumber \\
S_2 & =  {\phi}_1^{2} \sum_{\substack{a=1 \\ i=1,2}}^{N} (a-1) (B_a^i)^2 +  \phi_2^2 \sum_{\substack{a=1 \\ i=1,2}}^{N-1}  (N-a)  (B_{a+1}^i)^2 \,, \nonumber \\
S_3 & =  {\phi}_1^{2} \sum_{\substack{a=1 \\ i=1,2}}^{N} (a -1)  A_a^i  B_a^i +  \phi_2^2 \sum_{\substack{a=1 \\ i=1,2}}^{N-1}  (N-a) A_a^i  B_{a+1}^i \,.
\label{Scof}
\end{align}
The equation of motion for $\alpha$ and $\beta$ take the form
\be
\alpha S_1 - \beta S_3 =0 \,, \quad \beta S_2- \alpha S_3 = 0 \,,
\ee
which yields
\be
(S_1 S_2- S_3^2) \alpha = 0 \,.
\ee
Since $S_1 S_2 -S_3^2 \neq 0$, as one can see readily see by inspection (this is also verified using \emph{Mathematica} at several different choices of $N$), therefore, the only solution to the equations of motion is the trivial solution,
\be
\alpha(t) = 0 \,, \quad \beta(t) = 0 \,,
\ee
as we intended to show.

\subsection{Ansatz II}

In this case, we immediately see by inspection from \eqref{Heom2} and \eqref{Scof} that
\begin{align}
\Tr\abs{D_i Q_1}^2+\Tr\abs{D_i Q_2}^2 + \abs{D_i R_1}^2+\Tr\abs{D_i R_2}^2 =\alpha^2 T_1 +\beta^2 T_2 - 2\alpha\beta T_3 \,,
\end{align}
where
\begin{align}
	T_1 & = (q^2 +r^2) \bigg ( (N-1) (A_N^i)^2 + (N-2) \sum_{\substack{a=1 \\ i=1,2}}^{N-1} (A_a^i)^2 \bigg) \,, \nonumber \\
	T_2 & =  (q^2 +r^2) N  \sum_{\substack{a=2 \\ i=1,2}}^{N} (B_a^i)^2 \,, \nonumber \\
	T_3 & = (q^2 +r^2) \bigg( \sum_{\substack{a=1 \\ i=1,2}}^{N} (a -1)  A_a^i  B_a^i + \sum_{\substack{a=1 \\ i=1,2}}^{N}  (N-a) A_a^i  B_{a+1}^i \bigg ) \,.
\end{align}
with the equations of motion implying that 
\be
(T_1 T_2- T_3^2) \alpha = 0 \,.
\ee
Since, $T_1 T_2 -T_3^2 \neq 0$, by inspection,  the only solution to the equations of motion is, once again, the trivial solution $\alpha(t)= 0= \beta(t)$.

\section{Fixed points of the reduced Lagrangians and their stability} 
\label{AppB}

\subsection{Ansatz I}

Fixed points of a Hamiltonian system are defined as the stationary points of the phase space \cite{Baskan:2019qsb}. For the reduced dynamical system obtained using ansatz I, these points are given by the solutions of the equations
\begin{align}\label{fixed}
(\dot{\phi_1}, \dot{\phi_2}, \dot{p}_{\phi_1}, \dot{p}_{\phi_2} ) = (0,0,0,0) \,.
\end{align}
Combining \eqref{fixed} and the equations of motion given in \eqref{e.o.m} leads to four algebraic equations, two of which are trivially solved by  $(p_{\phi_1},p_{\phi_2}) \equiv(0,0)$. The remaining two give us the coupled algebraic equations, which are expressed as 
\begin{align}
&N^2(N-1)\left(\mu^2\phi_1+\frac{16\pi \mu }{k} \phi_1\phi_2^2+ \frac{16\pi^2}{k^2} \phi_1\phi_2^4 +\frac{32\pi^2}{k^2} \phi_1^3\phi_2^2 \right) = 0\,,\\
&N^2(N-1)\left(\mu^2{\phi}_2+\frac{16\pi \mu }{k} \phi_1^2{\phi}_2 + \frac{16\pi^2}{k^2} \phi_1^4{\phi}_2 +\frac{32\pi^2}{k^2} \phi_1^2{\phi}_2^3 \right) =0.
\label{fix1}
\end{align}
Fixed points may be determined by solving these equations.

Linear stability of the system may be inspected around a given fixed point, in order to determine whether it is a stable or unstable fixed point. A similar analysis was performed in Ref. \cite{Baskan:2019qsb} and we follow it in what follows. For simplicity, let us introduce the notation
\begin{align} 
(q_1,q_2,q_3,q_4)\equiv(\phi_1,\phi_2,p_{\phi_1},p_{\phi_2}) \,.
\end{align}
From $q_\alpha$ and $\dot{q}_\alpha$, we may form the Jacobian matrix
\begin{align}
J\equiv [J]_{\alpha\beta}&=\frac{\partial \dot{q}_\alpha}{\partial q_\beta} \,. 
\end{align}
The explicit form of $J$ is given as
\begin{align}
J=\begin{pmatrix}
0&0&\frac{1}{N^2(N-1)}&0\\
0&0&0&\frac{1}{N^2(N-1)}\\
J_{31}&J_{32}&0&0\\
J_{41}&J_{42}&0&0 
\end{pmatrix} \,,
\label{Jac1}
\end{align}
where 
\begin{align}
J_{31}&=-N^2(N-1)\left(\mu^2+\frac{96\pi }{k^2} \phi_1^2\phi_2^2 +\frac{16\pi^2}{k^2}\phi_2^4 +\frac{16\pi \mu}{k^2}\phi_2^2 \right) \,, \nonumber \\
J_{32}&=-N^2(N-1)\left(\frac{64 \pi^2 }{k^2} \phi_1^3\phi_2 + \frac{64 \pi^2 }{k^2} \phi_1 \phi_2^3 + \frac{32\pi \mu}{k} \phi_1 \phi_2 \right)\,, \nonumber \\
J_{41}&=-N^2(N-1)\left(\mu^2+\frac{96\pi }{k^2}\phi_1^2\phi_2^2 +\frac{16\pi^2}{k^2}\phi_1^4 +\frac{16\pi \mu}{k^2}\phi_1^2 \right) \,, \nonumber \\
J_{42}&=-N^2(N-1)\left(\frac{64 \pi^2 }{k^2}\phi_1^3\phi_2 +\frac{64 \pi^2 }{k^2}\phi_1 \phi_2^3 +\frac{32\pi \mu }{k} \phi_1\phi_2\right) \,.
\end{align}
Eigenvalues of $J$ allow us to determine characteristic of a given fixed point. As was summarized in Ref. \cite{Baskan:2019qsb}, a fixed point is unstable if the Jacobian has at least one real positive eigenvalue. It may be that all the nonvanishing eigenvalues can be purely imaginary. The latter case is a fixed point of borderline type for which the first order stability analysis is inconclusive and considerations beyond first order are necessary to decide on the characteristic of such a point.

At $\mu=1$, for $k>0$,  the only real solution of the system given in \eqref{fix1} is the trivial solution $(\phi_1,\phi_2)\equiv(0,0)$. Thus, the only fixed point of this Hamiltonian system is given as $(\phi_1,\phi_2,p_{\phi_1},p_{\phi_2}) \equiv(0,0,0,0)$ with vanishing energy; i.e., we have $E_F(0,0,0,0)=0$. We find that the eigenvalues of $J(0,0,0,0)$ are given as $\{\pm i,\pm i \}$. Thus, this fixed point is of borderline type. We will not perform a higher-order analysis for this fixed point.

For $k<0$, the fixed points of the system are given as 
\be
(\phi_1,\phi_2,p_{\phi_1},p_{\phi_2})=  \bigg \lbrace (0,0,0,0),\left(\pm (\mp) \frac{\sqrt{-k \mu }}{2\sqrt{\pi}},\pm\frac{\sqrt{-k \mu }}{2\sqrt{\pi}},0,0\right), 
\left(\pm (\mp)\frac{\sqrt{-k \mu }}{2\sqrt{3\pi}},\pm\frac{\sqrt{-k \mu }}{2\sqrt{3\pi}},0,0\right) \bigg \rbrace \,.
\ee
Eigenvalues of the Jacobian at these fixed points are given as
\begin{align}
J(0,0,0,0)& \rightarrow \{i \mu,i \mu ,-i \mu,-,i \mu\} \,, \nonumber \\
J\left(\pm (\mp) \frac{\sqrt{-k \mu}}{2\sqrt{\pi}},\pm\frac{\sqrt{- k \mu}}{2\sqrt{\pi}},0,0\right) & \rightarrow \{2i \mu, 2 i \mu, -2 i \mu ,-2 i \mu \}  \,, \nonumber \\
J\left(\pm (\mp)\frac{\sqrt{-k \mu}}{2\sqrt{3\pi}},\pm\frac{\sqrt{-k \mu}}{2\sqrt{3\pi}},0,0\right) & \rightarrow \left\{-\frac{2 \mu}{\sqrt{3}},\frac{2 \mu}{\sqrt{3}},-\frac{2}{3} i \sqrt{5} \mu,\frac{2}{3} i \sqrt{5} \mu \right\} \,.
\end{align}
Thus, the unstable fixed points are $\left(\pm (\mp)\frac{\sqrt{-k \mu}}{2\sqrt{3\pi}},\pm\frac{\sqrt{-k \mu}}{2\sqrt{3\pi}},0,0\right)$,  since their Jacobians have a positive eigenvalue, while the remaining are of borderline type. The energy of the system at the unstable fixed point is evaluated to be 
\be
E_F =  N^2 (N-1) \frac{5 | k \mu^3 |}{108 \pi } \,, \quad k  \mu < 0  \,. 
\ee

\subsection{Ansatz II}

Using the equations of motion given in \eqref{Heom2}, we find that the fixed points are determined by $(p_q, p_r) = (0,0)$ and the solutions of the equations  	
\begin{align}
	&\left(2\mu^2q+\frac{32\pi \mu}{k}q^3-\frac{48\pi^2}{k^2}q^3 r^2 -\frac{24\pi^2}{k^2}qr^4 + \frac{96\pi^2}{k^2}q^5\right)=0 \,, \nonumber\\ 
	&\left(2\mu^2r- \frac{32\pi \mu}{k}r^3 -\frac{48\pi^2}{k^2}q^2 r^3 -\frac{24\pi^2}{k^2}q^4 r+ \frac{96\pi^2}{k^2}r^5\right)=0 \,.
	\end{align}
We find that, for $k \mu > 0 $, they are given as
\begin{align}\label{fixedansatz2}
(q,r,p_q,p_r)=\{(0,\pm\frac{\sqrt{ k \mu}}{2\sqrt{\pi}},0,0),(0,\pm\frac{\sqrt{ k \mu}}{2\sqrt{3\pi}},0,0), (0,0,0,0)\} \,,
\end{align}
while for $k \mu < 0 $, they are
\begin{align}\label{fixedansatz3}
(q,r,p_q,p_r)=\{(\pm\frac{\sqrt{ - k \mu}}{2\sqrt{\pi}},0,0,0),(\pm\frac{\sqrt{- k \mu}}{2\sqrt{3\pi}},0,0,0), (0,0,0,0)\} \,.
\end{align}
The Jacobian matrix is again of the form given in \eqref{Jac1}, where now 
\begin{align}
	J_{31}&= - N^2(N-1) \left ( \frac{480 \pi^2 }{k^2}q^4 - \frac{144 \pi ^2 }{k^2}q^2 r^2 -\frac{24 \pi^2 }{k^2}r^4+\frac{96
		\pi  \mu }{k} q^2 +2 \mu ^2  \right )  \,, \nonumber \\
	J_{32}&= - N^2(N-1) \left ( \frac{96 \pi ^2 }{k^2}q^3 r +\frac{96 \pi ^2 }{k^2} q r^3 \right )       \,, \nonumber \\
	J_{41}&=- N^2(N-1)  \left ( \frac{96 \pi ^2 }{k^2}q^3 r +\frac{96 \pi ^2 }{k^2} q r^3 \right )        \,, \nonumber \\
	J_{42}&=- N^2(N-1)   \left (\frac{480 \pi ^2 }{k^2}r^4 -\frac{144 \pi ^2 }{k^2}q^2 r^2 -\frac{24 \pi ^2}{k^2} q^4 -\frac{96
		\pi  \mu  }{k}r^2 +2 \mu ^2 \right )    \,.
\end{align}
Eigenvalues of the Jacobian matrix at the fixed points are
\be
J(0,0,0,0) \rightarrow \left\{-i \sqrt{2} \mu ,-i \sqrt{2} \mu ,i \sqrt{2} \mu ,i \sqrt{2} \mu \right\} \,,
\ee
\begin{align}
J(0,\pm\frac{\sqrt{ k \mu}}{2\sqrt{\pi}},0,0) & \rightarrow \left\{-\frac{i \mu }{\sqrt{2}},\frac{i \mu }{\sqrt{2}},- 2 \sqrt{2} i \mu ,2 \sqrt{2} i \mu
\right\} \,, \nonumber \\ 
J(0,\pm\frac{\sqrt{k \mu}}{2\sqrt{3\pi}},0,0)& \rightarrow \left\{-2 \sqrt{\frac{2}{3}} \mu ,2 \sqrt{\frac{2}{3}} \mu ,-i \sqrt{\frac{11}{6}} \mu ,i
\sqrt{\frac{11}{6}} \mu \right\} \,,
\end{align}
\begin{align}
	J(\pm\frac{\sqrt{ - k \mu}}{2\sqrt{\pi}},0,0,0) & \rightarrow \left\{-\frac{i \mu }{\sqrt{2}},\frac{i \mu }{\sqrt{2}},- 2 \sqrt{2} i \mu ,2 \sqrt{2} i \mu
	\right\} \,, \nonumber \\ 
	J(\pm\frac{\sqrt{- k \mu}}{2\sqrt{3\pi}},0,0,0) & \rightarrow \left\{-2 \sqrt{\frac{2}{3}} \mu ,2 \sqrt{\frac{2}{3}} \mu ,-i \sqrt{\frac{11}{6}} \mu ,i
	\sqrt{\frac{11}{6}} \mu \right\} \,.
\end{align}
Therefore, for real values of $\mu$, we have either the set $(0,\pm\frac{\sqrt{k \mu}}{2\sqrt{3\pi}},0,0)$ or the set $(\pm\frac{\sqrt{- k \mu}}{2\sqrt{3\pi}},0,0,0)$ as unstable fixed points for $k \mu > 0$ and $k \mu < 0$, respectively, while the remaining fixed points are of borderline type.  The corresponding energy value at these points are
\be
E_F = N^2 (N-1) \frac{ |k \mu ^3|}{27 \pi } \,.
\ee
Let us finally observe that for purely imaginary values of $\mu$ (i.e., for tachyonic mass values) all these fixed points are of unstable type.

\end{document}